\documentclass[12pt,a4paper]{article}
\usepackage{float}
\usepackage{subfig}
\usepackage{amsmath}
\usepackage{xcolor}
\usepackage{ulem}
\usepackage{amssymb}
\usepackage{graphicx}
\usepackage{comment}
\usepackage{hyperref}
\usepackage[compat=1.0.0]{tikz-feynman}
\usepackage{csquotes}
\usepackage{multirow}
\usepackage[left=.45in,right=.45in,top=.6in,bottom=.6in,headheight=14.5pt]{geometry}
\usepackage{multirow}
\usepackage{pdflscape}
\usepackage{appendix}
\usepackage[left=.45in,right=.45in,top=.6in,bottom=.6in,headheight=14.5pt]{geometry}
\usepackage{fmtcount} 
\usepackage{array,multirow}
\newcolumntype{C}[1]{>{\centering\arraybackslash}m{#1}}
\usepackage{epsfig}
\usepackage{float}
\usepackage{geometry}
\usepackage{amsmath}
\usepackage{cite}
\usepackage{ctable}

\newgeometry{vmargin={10mm,20mm},hmargin={19mm,19mm}}

\begin{document}

\title{Low Scale Leptogenesis in Singlet-Triplet Scotogenic Model}

\author{Labh Singh$^{1}$\thanks{sainilabh5@gmail.com}, Devabrat Mahanta$^{2}$\thanks{deva.m2910@gmail.com}, and Surender Verma$^{1}$\thanks{s\_7verma@hpcu.ac.in}}

\date{%
$^{1}$Department of Physics and Astronomical Science, Central University of Himachal Pradesh, Dharamshala, India\\
$^{2}$Department of Physics, Abhayapuri College, Bongaigaon, Assam, India\\
}

\maketitle

\begin{abstract}
\noindent The scotogenic model presents an elegant and succinct framework for elucidating the origin of tiny neutrino masses within the framework of the Standard Model, employing radiative corrections within the domain of the dark sector. We investigate the possibility of achieving low-scale leptogenesis in the singlet-triplet scotogenic model (STSM), where dark matter mediates neutrino mass generation. We initially considered a scenario involving two moderately hierarchical heavy fermions, N and $\Sigma$, wherein the lepton asymmetry is generated by the out-of-equilibrium decay of both particles. Our analysis indicates that the scale of leptogenesis in this scenario is similar to that of standard thermal leptogenesis and is approximately $M_{N,\Sigma}\sim 10^{9}$ GeV, which is comparable to the Type-I seesaw case. Further, we consider the case with three heavy fermions ($N_1$, $N_2$, and $\Sigma$) with the hierarchy $M_{N_{1}} < M_{\Sigma} \ll M_{N_{2}}$, which yields the lower bound on heavy fermions up to 3.1 TeV, therefore significantly reduce the scale of the leptogenesis up to TeV scale. The only prerequisite is suppression in the $N_{1}$ and $\Sigma$ Yukawa couplings, which causes suppressed washout effects and a small active neutrino mass of about $10^{-5}$ eV. This brings about the fascinating insight that experiments aiming to measure the absolute neutrino mass scale can test low-scale leptogenesis in the scotogenic model. Further, the hyperchargeless scalar triplet $\Omega$ provides an additional contribution to mass of the $W$-boson explaining CDF-II result.

\end{abstract}
\section{Introduction}\label{intro}
\noindent Although the early Universe started with equal amounts of matter and antimatter, the present Universe contains an excess of matter over antimatter. To account for this asymmetry, a dynamical mechanism must be identified so that, after matter and antimatter annihilate, a surplus of matter remains, giving rise to the visible matter we presently observe. This surplus of baryons over antibaryons, is often expressed in term of baryon-to-photon ratio, $\eta_B$\cite{Workman:2022ynf,Planck:2018vyg}:
$$
\eta_B=\frac{n_B-n_{\bar{B}}}{n_\gamma}=6.1 \times 10^{-10},
$$
\noindent where the variables $n_{B}$, $n_{\bar{B}}$, and $n_{\gamma}$ correspond to the number densities of baryons, antibaryons and photons, respectively. In order to create this asymmetry dynamically, certain prerequisites known as ``Sakharov conditions" are required to exist \cite{Sakharov:1967dj}. These conditions are: (i) violation of baryon number (B) conservation, (ii) the existence of $\mathrm{C}$ and $\mathrm{CP}$ violation, and (iii) departure from thermal equilibrium during the early Universe. Although the SM falls short in satisfactorily fulfilling all these criteria, different theories beyond the Standard Model (BSM) have been proposed to explain observed baryon abundance in the Universe.

\noindent One of the mechanism among these hypotheses is leptogenesis \cite{Fukugita:1986hr} where the addition of extra massive Majorana neutrinos can decay into the SM particles in a way that meets all the aforementioned prerequisites for successful baryogenesis \cite{Weinberg:1979bt,Kolb:1979qa}.
Leptogenesis models based on seesaw scenarios provide promising framework for explaining small neutrino mass and baryon asymmetry of the Universe (BAU) within a single theoretical setting. In order to have successful leptogenesis, in Type-I and Type-II seesaw extensions, the heavy singlet fermion and scalar triplet, respectively, contribute to the overall lepton asymmetry which further can be converted to baryon asymmetry via sphaleron processes \cite{Luty:1992un,Plumacher:1996kc,Covi:1996wh,Antusch:2004xy,Buchmuller:2005eh,Antusch:2005tu,Davidson:2008bu,Chongdar:2021tgm,Kuzmin:1985mm}. The phenomena of leptogenesis in various extensions  of the SM has been discussed thoroughly in the literature \cite{Sarma:2020msa,Mahanta:2019gfe,Hugle:2018qbw,RickyDevi:2023fqd,Kashav:2021zir,Verma:2019uiu,Narendra:2018vfw,Baumholzer:2018sfb,Borah:2018uci,Huang:2018vcr,Alvarez:2023dzz,Mahapatra:2023dbr,Vatsyayan:2022rth,Mishra:2022egy,Albright:2003xb,Biswas:2023azl,Kashiwase:2012xd,Kashiwase:2013uy,Racker:2013lua,Cai:2017jrq}. The standard thermal leptogenesis based on Type-I seesaw model has a limitation of requiring very heavy Majorana fermions\cite{Buchmuller:2002rq, Giudice:2003jh,Buchmuller:2004nz,Davidson:2002qv}. In such models, for thermal leptogenesis to be successful, the lightest Majorana fermion need to have a minimum mass of around $10^{9}$ GeV known as the Davisson-Ibarra bound\cite{Davidson:2002qv}. By incorporating the flavor effects, this lower bound can, further, be brought down to $\approx10^{8}$ GeV\cite{Blanchet:2008pw} but it is still difficult to validate such a high-scale phenomenon in future collider experiments. This serves as motivation to seek alternative leptogenesis models that can produce the BAU at a lower energy scale without considering degenerate heavy fermions.

\noindent Keeping this motivation in mind, we revisit the singlet-triplet scotogenic model (STSM)\cite{Hirsch:2013ola,Rocha-Moran:2016enp,Fiaschi:2018rky,Choubey:2017yyn,Batra:2022org,Diaz:2016udz,Ding:2020vud,Restrepo:2019ilz,Avila:2019hhv,Belanger:2022gqc} which extends the original idea of the minimal scotogenic model presented in Ref. \cite{Ma:2006km} to make its phenomenology more viable and diverse. In this model, the SM is extended by a singlet ($N$) and a triplet ($\Sigma$) fermions, both odd under $Z_{2}$ symmetry with no hypercharge. The scalar sector of the SM is extended by a $Z_{2}$-even real triplet field ($\Omega$) and a $Z_{2}$-odd doublet field ($\eta$), with hypercharges 0 and 1/2, respectively. $\Omega$ facilitates the mixing between singlet and triplet fermionic fields, which contributes to neutrino mass at one loop level. Also, at the tree level, $\Omega$ introduces additional contribution to the $W$-mass, as observed in CDF-II\cite{CDF:2022hxs}, and has been explored in Ref. \cite{Batra:2022org} within the context of STSM.

\noindent The low-scale leptogenesis in the minimal scotogenic model has been discussed in Refs. \cite{Mahanta:2019gfe,Hugle:2018qbw} where out-of-equilibrium decay of the lightest singlet fermion contribute to the final lepton asymmetry. In the present work, we have investigated the possibility of low-scale leptogenesis in STSM. We initially consider a scenario involving two moderately degenerate heavy fermions ($M_{N}\lesssim M_{\Sigma}$) wherein lepton asymmetry is generated by the out-of-equilibrium decay of both singlet and triplet fermions. In this case, the Yukawa couplings ($Y_N,Y_\Sigma$) are large enough keeping leptogenesis in the strong washout regime. This leads to high scale of leptogenesis wherein $M_{(N,\Sigma)}\geq 10^{9}$ GeV. In the second scenario, we have added two singlet right-handed fermions ($N_{1}$ and $N_{2}$) along with a triplet fermion $\Sigma$ in order to lower the scale of leptogenesis. Also, we consider a specific mass hierarchy among the heavy fermions ( $M_{N_{1}} \lesssim M_{\Sigma} \ll M_{N_{2}}$) so that decay of both $N_1$ and $\Sigma$ contribute to the final lepton asymmetry . The Yukawa couplings of $N_1$ exhibit a dependence on the lightest neutrino mass, $m_{1}$. As $m_{1}$ approaches vanishingly small values, the Yukawa coupling associated with the lightest fermion, $N_{1}$, concurrently diminishes to an extent where a weak washout leptogenesis scenario ensues. For $m_{1}$ $\lesssim 10^{-5}$ eV, the effects of washouts become negligible. This significantly reduces the leptogenesis scale to the $\rm TeV$ scale.

\noindent Furthermore, we have studied the implication of the model for relic density of the dark matter (DM). There are two possible DM candidates in the model (i) fermionic DM i.e. mixed state of singlet and triplet fermion, and (ii) the lightest component of the inert scalar doublet ($\eta$). In the fermionic DM case, the relic density requires larger Yukawa couplings which results in large washout effects in leptogenesis\cite{Ma:2006fn}. However, in the scalar dark matter scenario, the relic density depends on the gauge interactions of the $\eta$. Therefore, realizing thermal leptogenesis with a specific mass spectrum is difficult in the former scenario. Also, it is important to note that while dealing with fermionic dark matter, the significantly large Yukawa couplings can give sizeable contribution to lepton flavor violations\cite{Adulpravitchai:2009gi,Toma:2013zsa,Vicente:2014wga}. Thus, we investigate the phenomenology of scalar DM where we chose real component ($\eta^{R}$) of the neutral part of $\eta$ as DM candidate.

\noindent The structure of the paper is as follows: In Section \ref{model}, we begin by introducing the STSM and providing an overview of its essential characteristics. Section \ref{wmass1} delves into the $W$-boson mass anomaly observed in the CDF-II results. Section \ref{dm} is devoted to the phenomenology of the scalar DM. In Section \ref{lepto}, we formulate all necessary expressions for studying leptogenesis in the STSM. Considering the scenario with two heavy fermions, the analysis shows that this scenario offers no advantage over the standard thermal leptogenesis in Type-I seesaw framework (Section \ref{lepto1}). We extend our model in Section \ref{lepto2} by introducing another right-handed singlet fermion to lower down the scale of leptogenesis. Section \ref{result} will briefly present numerical analysis and discussion of leptogenesis scenario. In Section \ref{summary}, we present conclusions of the work reported in this paper.

\section{Model and Formalism}\label{model}

The assignments of fields within the STSM, subject to the $SU(2)_L \times U(1)_Y \times Z_2$ symmetry, are presented in Table \ref{tab1}. The relevant invariant Yukawa Lagrangian is given by
\begin{eqnarray}
\mathcal{L}=Y^{\alpha \beta}_{L} \overline{L_{\alpha}}\phi \ell_{\beta}+Y^{\alpha}_{N}\overline{L_{\alpha}}i \sigma_{2} \eta N+Y^{\alpha}_{\Sigma}\overline{L_{\alpha}}C \Sigma^{\dagger}i \sigma_{2} \eta+Y_{\Omega} Tr(\overline{\Sigma}\Omega)N+ \frac{1}{2}M_{N}\overline{N^{c}}N+\frac{1}{2}M_{\Sigma}Tr(\overline{\Sigma^{c}}\Sigma)+h.c. ,
\end{eqnarray}
where $\alpha$, $\beta$=$1,2,3$ denotes the three flavors.
The $S U(2)_L$ doublet fields $(\phi, \eta)$ and $S U(2)_L$ triplet fields $(\Sigma, \Omega)$ can be expressed as
$$
\phi=\left(\begin{array}{c}
\phi^{+} \\
\phi^0
\end{array}\right), \eta=\left(\begin{array}{c}
\eta^{+} \\
\eta^0
\end{array}\right), \Sigma=\left(\begin{array}{cc}
\frac{\Sigma^0}{\sqrt{2}} & \Sigma^{+} \\
\Sigma^{-} & -\frac{\Sigma^0}{\sqrt{2}}
\end{array}\right) \text { and } \quad \Omega=\left(\begin{array}{cc}
\frac{\Omega^0}{\sqrt{2}} & \Omega^{+} \\
\Omega^{-} & -\frac{\Omega^0}{\sqrt{2}}
\end{array}\right).
$$
The invariant scalar potential of the model is given by
\begin{eqnarray}
\nonumber
V&=&-m_{\phi}^2 \phi^{\dagger} \phi+m_\eta^2 \eta^{\dagger} \eta+\frac{\lambda_1}{2}\left(\phi^{\dagger} \phi\right)^2+\frac{\lambda_2}{2}\left(\eta^{\dagger} \eta\right)^2+\frac{m_{\Omega}^2}{2} \operatorname{Tr}\left(\Omega^{\dagger} \Omega\right)+\lambda_3\left(\phi^{\dagger} \phi\right)\left(\eta^{\dagger} \eta\right)+\lambda_4\left(\phi^{\dagger} \eta\right)\left(\eta^{\dagger} \phi\right)\\
\nonumber
&+&\frac{\lambda_5}{2}\left[\left(\phi^{\dagger} \eta\right)^2+h.c.\right]+\frac{\lambda^\eta}{2}\left(\eta^{\dagger} \eta\right) \operatorname{Tr}\left(\Omega^{\dagger} \Omega\right)+\frac{\lambda_1^{\Omega}}{2}\left(\phi^{\dagger} \phi\right) \operatorname{Tr}\left(\Omega^{\dagger} \Omega\right)+\frac{\lambda_2^{\Omega}}{4} \operatorname{Tr}\left(\Omega^{\dagger} \Omega\right)^2\\
&+&\mu_1 \phi^{\dagger} \Omega \phi+\mu_2 \eta^{\dagger} \Omega \eta.
\end{eqnarray}

\begin{table}[h]
    \centering
    \begin{tabular}{ccc}
    \hline\hline
    Particle Content & Generations & Symmetry\\
    &&$(SU(2)_{L}\times U(1)_{Y}\times Z_{2})$\\
    \hline\hline
    $L$ & 3 & $( 2, -\frac{1}{2}, + )$\\
    $\ell$ & 3 & $( 1, -1, + )$\\
    $\phi$ & 1 &$( 2, \frac{1}{2}, + )$\\
    $N$ & 1 &$( 1, 0, - )$\\
    $\Sigma$ & 1 &  $( 3, 0, - )$\\
    $\Omega$ & 1 & $( 3, 0, + )$\\
    $\eta$ & 1 & $( 2, \frac{1}{2},- )$ \\
    \hline\hline
    \end{tabular}
    \caption{Particle content, generations, and symmetry assignments in the STSM under $SU(2)_{L}\times U(1)_{Y}\times Z_{2}$ gauge symmetry.}
    \label{tab1}
\end{table}
\noindent In order to maintain the perturbativity of the theory, it is required that all couplings ($\lambda_{i}$) remain less than or equal to one\cite{Merle:2016scw,Kannike:2012pe}. Due to the conservation of $\mathbb{Z}_2$ symmetry, the $\eta$ field does not acquire a vacuum expectation value (VEV). The spontaneous electroweak symmetry breaking is triggered primarily by the neutral components of the $\phi$ and $\Omega$ fields,
$$
\left\langle\phi^0\right\rangle=\frac{v_{\phi}}{\sqrt{2}}, \quad\left\langle\Omega^0\right\rangle=v_{\Omega}.
$$
The scalar spectrum is categorized into two distinct sets: the $\mathbb{Z}_2$-even scalars, namely $\phi^0$, $\Omega^0$, $\Omega^{\pm}$, and $\phi^{\pm}$, and the $\mathbb{Z}_2$-odd scalars, namely $\eta^0$ and $\eta^{\pm}$. Among these, $\eta^0$ has garnered significant attention as a promising dark matter candidate, with extensive studies documented in the literature\cite{Dolle:2009fn,Deshpande:1977rw,Barbieri:2006dq,LopezHonorez:2006gr,LopezHonorez:2010eeh,Garcia-Cely:2015khw,Queiroz:2015utg}. Within this framework, the neutral scalars $\phi^0$ and $\Omega^0$ undergo mixing through a $2 \times 2$ matrix that can be parameterized by the angle $\Theta$, such that
\begin{eqnarray}
\left(\begin{array}{l}
H_{1} \\
H_{2}
\end{array}\right)=\left(\begin{array}{cc}
\cos \Theta & \sin \Theta \\
-\sin \Theta & \cos \Theta
\end{array}\right)\left(\begin{array}{l}
\phi^0 \\
\Omega^0
\end{array}\right),
\end{eqnarray}
where
\begin{eqnarray}
\tan (2 \Theta)=\frac{4 v_{\Omega} v_\phi\left(\sqrt{2} \mu_1-2 \lambda_1^{\Omega} v_{\Omega}\right)}{8 \lambda_2^{\Omega} v_{\Omega}^3-4 \lambda_1 v_{\Omega} v_\phi^2+\sqrt{2} \mu_1 v_\phi^2} .
\end{eqnarray}
In this theory, the lightest $Z_2$-even scalar, $H_{1}$, is identified as the SM Higgs boson, while the heavier one, $H_{2}$, remains a proposed new scalar Higgs boson. Similarly, the charged scalars $\phi^{\pm}$ and $\Omega^{\pm}$ will undergo mixing described by a $2 \times 2$ matrix given by
\begin{eqnarray}
\left(\begin{array}{l}
H_{1}^{ \pm} \\
H_{2}^{ \pm}
\end{array}\right)=\left(\begin{array}{cc}
\cos \delta & \sin \delta \\
-\sin \delta & \cos \delta
\end{array}\right)\left(\begin{array}{c}
\phi^{ \pm} \\
\Omega^{ \pm}
\end{array}\right),
\end{eqnarray}
with
\begin{eqnarray}
\tan (2 \delta)=-\frac{4 v_{\Omega} v_\phi}{v_\phi^2-4 v_{\Omega}^2}.
\end{eqnarray}
It is to be noted that the lightest charged scalar, which is represented by $H_{1}^{\pm}$, is the Goldstone boson. Also, the theory introduces a novel charged scalar ($H_{2}^{\pm}$) field. Further, the masses of the $Z_2$-odd scalars $\eta^0$ and $\eta^{\pm}$ are given by
\begin{eqnarray}
& m_{\eta^{ \pm}}^2=m_\eta^2+\frac{1}{2} \lambda_3 v_\phi^2+\frac{1}{2} \lambda^\eta v_{\Omega}^2+\frac{1}{\sqrt{2}} v_{\Omega} \mu_2, \\
& m_{\eta^R}^2=m_\eta^2+\frac{1}{2} \lambda_3 v_\phi^2+\frac{1}{2}\left(\lambda_4+\lambda_5\right) v_\phi^2-\frac{1}{\sqrt{2}} v_{\Omega} \mu_2 +\frac{1}{2} \lambda^\eta v_{\Omega}^2, \\
& m_{\eta^I}^2=m_\eta^2+\frac{1}{2} \lambda_3 v_\phi^2+\frac{1}{2}\left(\lambda_4-\lambda_5\right) v_\phi^2-\frac{1}{\sqrt{2}} v_{\Omega} \mu_2 +\frac{1}{2} \lambda^\eta v_{\Omega}^2.
\end{eqnarray}
Within the fermionic sector, the fields $\Sigma^0$ and $N$, which possess $\mathbb{Z}_2$-odd properties, undergo mixing governed by Yukawa coupling $Y_{\Omega}$ and occur in presence of the non-zero vacuum expectation value, $v_{\Omega}$. In the basis of $\left(\Sigma^0, N\right)$, the Majorana mass matrix is given by
\begin{eqnarray}
M_\chi=\left(\begin{array}{cc}
M_{\Sigma} & v_{\Omega} Y_{\Omega} \\
v_{\Omega} Y_{\Omega} & M_N
\end{array}\right),
\end{eqnarray}
which is diagonalized by $2 \times 2$ matrix $V(\alpha)$ as
\begin{eqnarray}
\left(\begin{array}{l}
\chi_1^0 \\
\chi_2^0
\end{array}\right)=V(\alpha)\left(\begin{array}{c}
\Sigma^0 \\
N
\end{array}\right),
\end{eqnarray}
where $$V(\alpha)=\left(\begin{array}{cc}
\cos \alpha & \sin \alpha \\
-\sin \alpha & \cos \alpha
\end{array}\right).$$
Consequently, masses of $\chi^{ \pm}$ and $\chi_i^0$ eigenstates at tree-level are
\begin{eqnarray}
& m_{\chi^{ \pm}}=M_{\Sigma}, \\
& m_{\chi_{1,2}^0}=\frac{1}{2}\left(M_{\Sigma}+M_N\mp\sqrt{\left(M_{\Sigma}-M_N\right)^2+4\left(v_{\Omega} Y_{\Omega} \right)^2}\right),
\end{eqnarray}
where the mixing angle $\alpha$ satisfy the relation
\begin{eqnarray}
\tan{(2\alpha)}=\frac{2 v_{\Omega} Y_{\Omega}}{M_{\Sigma}-M_{N}}.
\end{eqnarray}

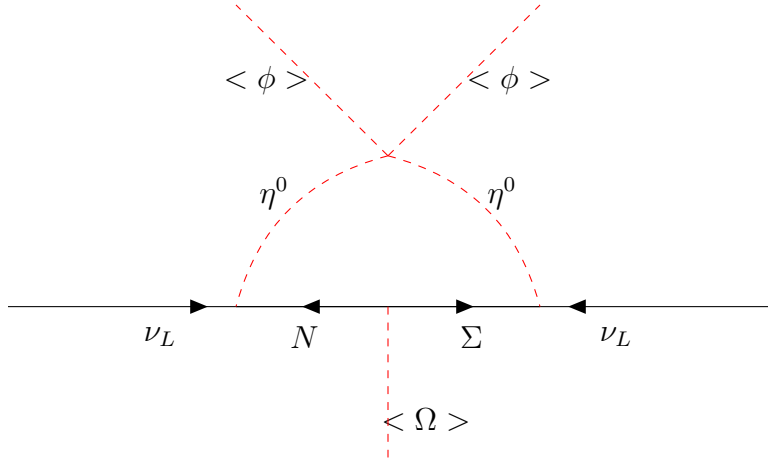
\begin{figure}[t]
    \centering
   \begin{tikzpicture}
\begin{feynman}
\vertex at (2,0) (i1);
\vertex at (-2,0) (i2);
\vertex at (0,0) (a);
\vertex at (0, 2) (d);
\vertex at (0,-2) (e);
\vertex at (5,0) (b);
\vertex at (-5,0) (c);
\vertex at (1.1,-0.4) () {\(\Sigma\)};
\vertex at (-1.1,-0.4) () {\(N\)};
\vertex at (0.5,-1.5) () {\(<\Omega>\)};
\vertex at (2,4) (f);
\vertex at (-2,4) (g);
\vertex at (1.6,3) () {\(<\phi>\)};
\vertex at (-1.6,3) () {\(<\phi>\)};
\vertex at (1.5,1.5) () {\(\eta^{0}\)};
\vertex at (-1.5,1.5) () {\(\eta^{0}\)};
\vertex at (-3,-0.4) () {\(\nu_{L}\)};
\vertex at (3,-0.4) () {\(\nu_{L}\)};
\diagram*{
(a) -- [fermion] (i1), (a) -- [fermion] (i2),
(b) -- [fermion] (a), (c) -- [fermion] (a),
(d) -- [red, scalar, bend left] (i1), (d) -- [red, scalar,bend right]  (i2),
(a) -- [red, scalar] (e), (d) -- [red, scalar] (f),
 (d) -- [red, scalar] (g), 
};
\end{feynman}
\end{tikzpicture}
    \caption{Feynman diagram used to generate neutrino masses at one-loop in STSM.}
    \label{fig:g2}
\end{figure}
\noindent In the model under investigation, the small neutrino masses are generated radiatively at the one-loop level as shown in Fig. \ref{fig:g2}. The real triplet scalar $\Omega$ whose VEV accounts for additional contribution to the $W$-boson mass, plays a vital role in neutrino mass generation as coupling $Y_{\Omega}$ generates a mixing between $N$ and $\Sigma$ fermions. At one-loop level, the neutrino mass matrix can be written as
\begin{eqnarray}
\left(\mathcal{M}_\nu\right)_{\alpha \beta} & =\sum_{i=1}^2 \frac{h_{\alpha i} h_{\beta i} m_{\chi_i^0}}{2(4 \pi)^2}\left[\frac{m_{\eta^R}^2 \ln \left(\frac{m_{\chi_i^0}^2}{m_{\eta^R}^2}\right)}{m_{\chi_i^0}^2-m_{\eta^R}^2}-\frac{m_{\eta^I}^2 \ln \left(\frac{m_{\chi_i^0}^2}{m_{\eta^I}^2}\right)}{m_{\chi_i^0}^2-m_{\eta^I}^2}\right], \\
&= \sum_{i=1}^{2} \frac{h_{\alpha i} h_{\beta i} m_{\chi_i^0}}{2(4 \pi)^2}[L_{i}(m_{\eta^R}^2)-L_{i}(m_{\eta^I}^2)],
\end{eqnarray}
which can be further written as
\begin{eqnarray}
=\sum_{i=1}^2 h_{\alpha i} \Lambda_i\left(h^T\right)_{i \beta}=\left(h \Lambda h^T\right)_{\alpha \beta}
\end{eqnarray}
where $h$ and $\Lambda$ matrices are given by
$$
h=\left(\begin{array}{cc}
Y_{\Sigma}^1 &  Y_N^1 \\
Y_{\Sigma}^2 &  Y_N^2 \\
Y_{\Sigma}^3 &  Y_N^3
\end{array}\right) V^T(\alpha), \quad \Lambda=\left(\begin{array}{cc}
\Lambda_1 & 0 \\
0 & \Lambda_2
\end{array}\right),
$$

with
\begin{eqnarray}
\Lambda_i=\frac{m_{\chi_i^0}}{2(4 \pi)^2}\left[\frac{m_{\eta^R}^2 \ln \left(\frac{m_{\chi_i^0}^2}{m_{\eta^R}^2}\right)}{m_{\chi_i^0}^2-m_{\eta^R}^2}-\frac{m_{\eta^I}^2 \ln \left(\frac{m_{\chi_i^0}^2}{m_{\eta^I}^2}\right)}{m_{\chi_i^0}^2-m_{\eta^I}^2}\right] .
\label{18}
\end{eqnarray}
In the scenario where the masses of $\eta^R$ and $\eta^I$ are equal, \textit{i.e.}, $m_{\eta^R}=m_{\eta^I}$, the resulting neutrino masses are zero, which leads to the condition $\lambda_5=0$, so the lepton number is conserved. One convenient way to represent the Yukawa couplings, $h_{\alpha i}$, is by employing the Casas-Ibarra parametrization\cite{Casas:2001sr,Ibarra:2003up} which turns out to be
\begin{eqnarray}
h=U^* \sqrt{\widetilde{M}} R \sqrt{\Lambda}^{-1},
\label{eq:CI}
\end{eqnarray}
where $U$ is Pontecorvo–Maki–Nakagawa–Sakata (PMNS) matrix, $\widetilde{M}=\operatorname{diag}\left(m_1, m_2, m_3\right)$ with $m_i$ are the neutrino physical masses and $\Lambda$ is given by Eqn. (\ref{18}). Additionally, the matrix $R$, which is a $3 \times 2$ orthogonal matrix satisfying $R R^T=\mathbb{I}_{3 \times 3}$, can be expressed as,
$$
\begin{aligned}
& R=\left(\begin{array}{cc}
0 & 0 \\
\cos \theta & \sin \theta \\
-\sin \theta & \cos \theta
\end{array}\right),
\end{aligned}  \hspace{0.5cm}
$$
where $\theta$ is a complex angle, and this description is applicable in the case of normal hierarchy only\cite{Fiaschi:2018rky,Choubey:2017yyn,Batra:2022org}.

\section{W-boson Mass Anomaly}\label{wmass1}
It is important to note that the pseudo-scalar $Z_{2}$-even portion solely comprises the imaginary component of the neutral part of $\phi$ and does not include any contributions from $\eta$ or $\Omega$. Therefore, $Z$-boson gets no contribution from $\Omega$. The inclusion of the real scalar triplet, $\Omega$, with zero hypercharge leads to the contribution in the mass of the $W$-boson ($M_{W}$). When the field $\Omega$ undergoes spontaneous symmetry breaking by acquiring a non-zero VEV, it imparts an extra contribution to the mass of the $W$-boson. This supplementary contribution to the mass of the $W$-boson serves as a potential explanation for the experimental findings observed by the CDF-II collaboration \cite{CDF:2022hxs}.
\begin{figure}
    \centering
    \includegraphics[width=13cm,height=8cm]{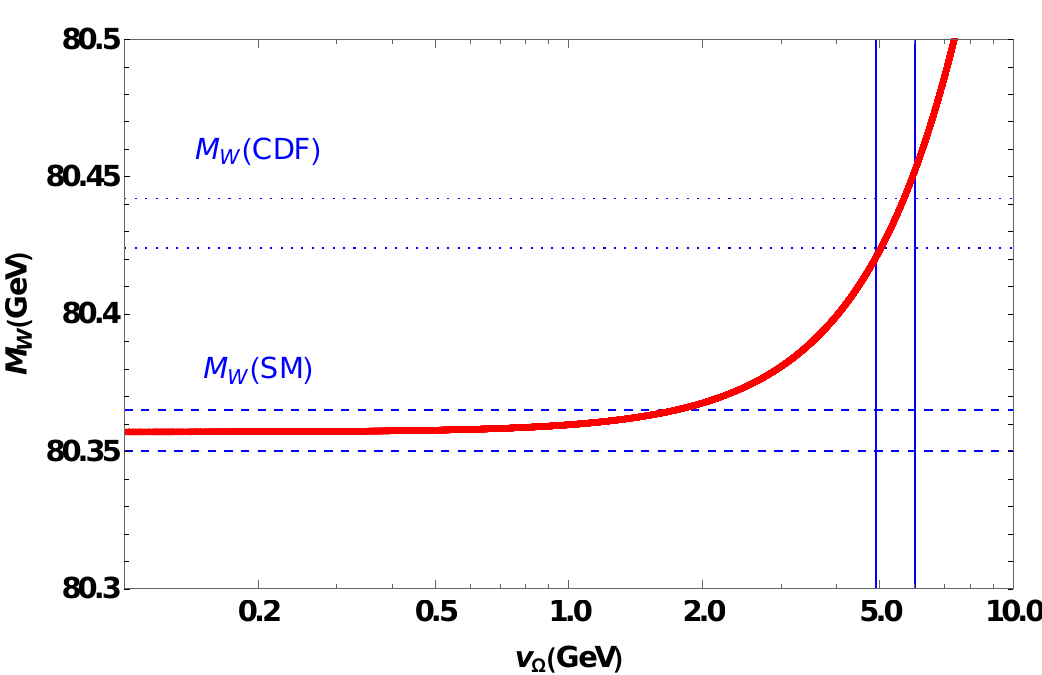}
    \caption{The plot illustrating the correlation between the $M_{W}$ and $v_{\Omega}$. The horizontal dashed lines represent the SM predictions, the horizontal dotted lines indicate the 1-$\sigma$ limit based on the CDF-II measurement, and the vertical solid lines signify the range of $v_{\Omega}$ values that match with the CDF-II outcome.}
    \label{wmass}
\end{figure}
The pertinent components within the Lagrangian that pertain to the mass of the $W$-boson, are 
\begin{eqnarray}\nonumber
\mathcal{L'}=\left(D_{\mu}\phi\right)^{\dagger} \left(D^{\mu}\phi\right)+Tr[\left(D_{\mu}\Omega\right)^{\dagger} \left(D^{\mu}\Omega\right)],   
\end{eqnarray}
where the covariant derivative is defined as
\begin{eqnarray}
D_{\mu}=\partial_{\mu}+ig\frac{\sigma_{a}}{2}W^{a}_{\mu}+ig'\frac{Y}{2}B_{\mu}.
\end{eqnarray}
The coupling constants $g$ and $g'$ correspond to the gauge symmetries $SU(2)_L$ and $U(1)_Y$, respectively. \\
Considering the non-zero value of $v_{\Omega}$ allows us to deduce the masses of the gauge bosons resulting from the contribution of $\Omega$. The expressions for the masses of the $W$ and $Z$ bosons are as
\begin{eqnarray}
M_{W}^{2}=(v_{\phi}^{2}+4v_{\Omega}^{2})\frac{g^{2}}{4}\hspace{0.3cm} \text{and} \hspace{0.3cm} M_{Z}^{2}=(g^{2}+g'^{2})\frac{v_{\phi}}{4}.
\end{eqnarray}
Since $\Omega$ lacks hypercharge, it remains uncoupled to the $Z$-boson, thus exerting no impact on its mass. The mass of the $W$-boson is now recognized to be influenced by scalar triplet's VEV $v_{\Omega}$, prompting the utilization of the latest CDF-II measurement to establish an upper limit on $v_{\Omega}$. The measured value of $M_{W}$ obtained by the CDF-II experiment can be expressed as follows
\begin{eqnarray}
(M_{\text{W}}^{2})^{\text{CDF}}-(M_{\text{W}}^{2})^{\text{SM}}=g^{2}v_{\Omega}^{2},
\label{23}
\end{eqnarray}
where $(M_{\text{W}}^{2})^{\text{SM}}$=$\frac{g^{2}v_{\phi}^{2}}{4}$ is the mass of $W$-boson predicted by SM and value of $g$= 0.6517\cite{ParticleDataGroup:2020ssz}.\\
Based on the latest CDF-II outcome and prediction of the SM, it becomes imperative to ensure, with in the framework, that the impact on the observed value of $M_{W}$ arises exclusively from the VEV of the real scalar triplet $\Omega$. 
Hence, the constraint delineated in Eqn. (\ref{23}) can be expressed as a bound on $v_{\Omega}$ which is given by
\begin{eqnarray}
4.8 \hspace{0.1cm} \text{GeV} \leq v_{\Omega} \leq 6 \hspace{0.1cm} \text{GeV},
\label{24}
\end{eqnarray}
and represented by the vertical band in Fig. \ref{wmass}. The CDF-II outcome and prediction by the SM are indicated by dotted and dashed horizontal lines, respectively. It is to be noted that the range of $v_\Omega$ obtained in Eqn. (\ref{24}) is in consonance with that of reported in Ref.\cite{Batra:2022org}.

\section{Phenomenology of Scalar Dark Matter}\label{dm}

\begin{table}[t]
\centering
\begin{tabular}{cc}
\hline\hline Parameter & Range \\
\hline\hline$M_{\Sigma}, M_N$ & {$\left[5\times10^3, 10^{12}\right] \mathrm{GeV}$} \\$m_\eta^2$ & {$\left[10^2, 10^{10}\right] \mathrm{GeV}$} \\
$\mu_i$ & {$[10,10^5] \mathrm{GeV}$} \\
$v_{\Omega}$ & {$[4.8,6] \mathrm{GeV}$} \\
$\lambda_5$ & {$\left[-1,-10^{-5}\right]$} \\
$\lambda_4$ &  {$\left[-1,-10^{-5}\right]$} \\
$\lambda_{2,3}$ & {$\left[10^{-5}, 10^{-1}\right]$} \\
$Y_{\Omega}$ & {$\left[10^{-5}, 10^{-1}\right]$} \\
$\lambda_i^{\Omega}$ & {$\left[10^{-5}, 10^{-1}\right]$} \\
$\lambda^\eta$ & {$\left[10^{-5}, 10^{-1}\right]$} \\
\hline\hline
\end{tabular}
\caption{\label{tab2}Input parameter's range used for the numerical analysis.}
\end{table}

\begin{figure}[t]
    \centering
    \includegraphics[scale=0.65]{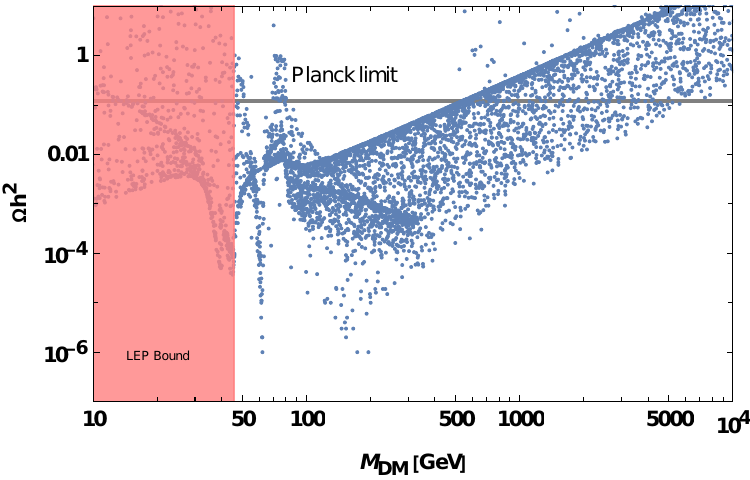}
    \includegraphics[scale=0.65]{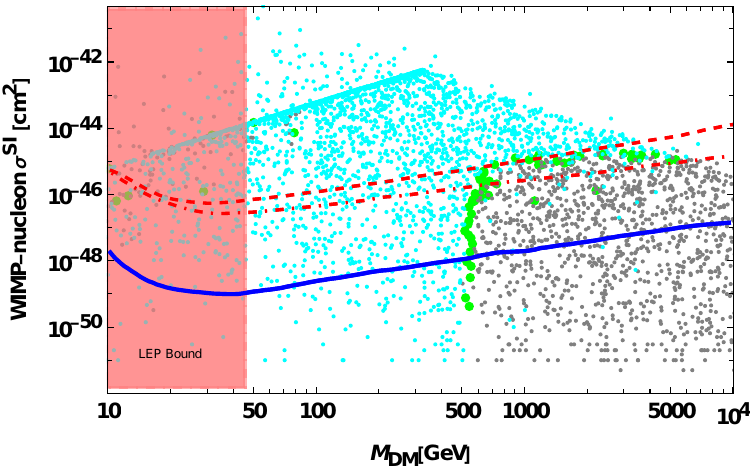}
    \caption{\textbf{Left Panel:} Correlation between DM Relic density and DM ($\eta^{R}$) mass, $M_{DM}$ for input values, as shown in Table \ref{tab2}. The grey horizontal band represents the 3$\sigma$ experimental range of $\Omega h^2$ coming from Planck observation\cite{Planck:2015fie,Planck:2018vyg}. LEP rule-out region with $m_{\eta}^R < \frac{m_Z}{2}$ and $m_{\eta^+} \lesssim 70 \, \text{GeV}$, is represented by the shaded region\cite{Belyaev:2016lok}. \textbf{Right Panel:} Correlation between Spin-independent DM-nucleon cross section vs $M_{DM}$. The green points signify values within the 3$\sigma$ range of $\Omega h^2$, while grey and cyan color points represent over and under-abundance, respectively.}
    \label{dm_dd}
\end{figure}

The model allows for the possibilities of both fermionic and bosonic DM candidates. In so far, fermionic DM is concerned, the lighter neutral eigenstates $\chi_{1}^{0}$ and $\chi_{2}^{0}$ are good candidates while for bosonic DM the lightest particle amongst the components of $\eta$ may serves as the potential DM candidate. Within the premise that the masses of $N$ and $\Sigma$ are sufficiently large, the preservation of $Z_{2}$ symmetry designates lighter of $\eta^{R}$ and $\eta^{I}$, as the promising candidate for DM. In our investigation, we contemplate $\eta^{R}$ as the preferred DM candidate, given the condition of $\lambda_{5} < 0$.
Alternatively, in the scenario of $\lambda_{5} >$ 0, $\eta^{I}$ would be considered the potential DM candidate. We implement the model in SARAH\cite{Staub:2015kfa} and subsequently employed in SPheno\cite{Porod:2011nf} to effectively compute all relevant mass matrices and vertices. Further, micrOMEGAs-5.0.8\cite{Belanger:2014vza} served as a critical tool in solving the Boltzmann equations and evaluating the thermal characteristics of the relic density of WIMP DM and calculating the cross-section for direct-detection (DD) of DM particle. In the left panel of Fig. \ref{dm_dd}, we depict the projected DM relic density ($\Omega h^{2}$) as a function of the DM mass ($M_{DM}$). The solid grey line represents the 3$\sigma$ range of relic density (0.1126$\leq\Omega h^2\leq0.1246$) inferred from data obtained by the Planck satellite\cite{Planck:2015fie,Planck:2018vyg}. We performed a numerical scan of the input parameters within the defined intervals outlined in Table \ref{tab2}. 
In order to ensure a mass of $125$ GeV for the lighter neutral scalar, $H_{1}$, the parameter $\lambda_{1}$ is adjusted accordingly.

\begin{figure}[t]
    \centering
    \includegraphics[width=11cm,height=7cm]{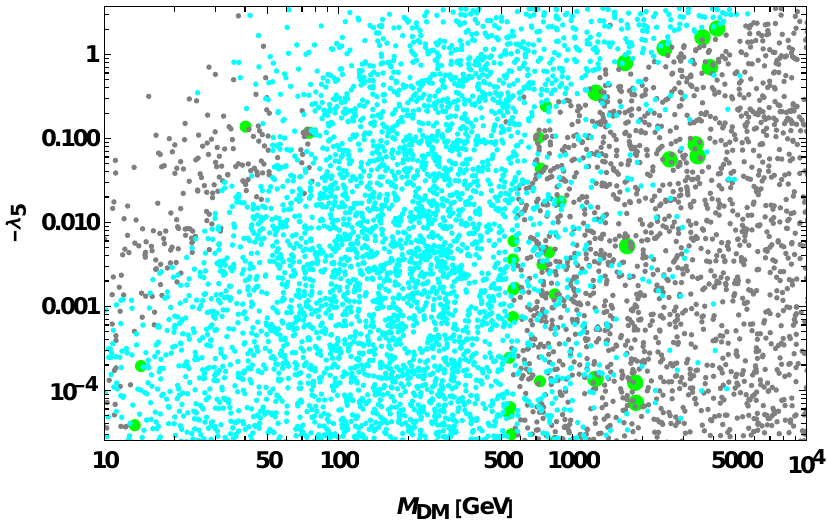}
    \caption{Variation in the $-\lambda_5$ as function of $M_{DM}$. The color scheme is similar to the right panel of Fig. \ref{dm_dd}}
    \label{dm_lambda5}
\end{figure}
\noindent The distinct dips observed can be understood by examining the annihilation and coannihilation processes, which are detailed in Appendix \ref{dmchanels}. The initial trough observed at $M_{\text{DM}}\sim \frac{M_{Z}}{2}$ arises due to annihilation and coannihilation processes facilitated by the exchange of the $Z$-boson in the $s$-channel. The subsequent dip, located at $M_{\text{DM}}\sim 60$ GeV, is linked to annihilation processes mediated through the $s$-channel exchange of $H_{1}$. In comparison, annihilation mediated by the exchange of $H_{1}$ is more efficient than that involving the $Z$-boson exchange, primarily due to the momentum suppression that hampers the latter. As the mass of $\eta^{R}$ increases, the quartic interactions with gauge bosons begin to play a significant role. For instance, the third dip in $\Omega h^2$ is brought on by the annihilation of DM candidate $\eta^R\rightarrow W^{+}W^{-},ZZ$ $via$ quartic couplings for $M_{\text{DM}}\geq$ 80 GeV. Moreover, for $M_{\text{DM}}\geq$ 120 GeV and $m_{\eta^{R}}\geq$ $m_{t}$ GeV, $\eta^{R}$ has the capability to undergo annihilation processes yielding two Higgs bosons ($H_{1}H_{1}$) and top-antitop quark pairs ($t\bar{t}$), respectively. Also, it is noted that, beyond 120 GeV, as $M_{\text{DM}}$ increases, the relic density rises. This phenomenon is attributed to the suppression of the annihilation cross-section, which decreases as $\sim\frac{1}{M_{\text{DM}}}$. Additionally, it is crucial to highlight that in parameter space regions characterized by a small $\lambda_{5}$, coannihilation processes involving both $\eta^{I}$ and $\eta^{\pm}$ can take place. It is evident from Fig. \ref{dm_dd} that bound on relic density of DM is satisfied for essentially one value of DM mass while keeping other parameters fixed.\\
Let us now explore the possibilities for DD of $\eta^R$. The interaction cross-section of DM and nucleons at the tree-level is governed by $\phi$ and $Z$ portals. Due to non-zero hypercharge of the $\eta$ doublet, the DM-nucleon interaction through the $Z$-boson can, in principle, surpass the latest constraints set by DD experiments. However, $\lambda_5$ introduces a mass difference between the CP-odd counterparts, $\eta^I$ and $\eta^R$. Consequently, the interaction involving the $Z$-boson is either prevented due to kinematic constraints or leads to inelastic scattering. Thus, the dominant contribution to the DM-nucleon interaction occurs $via$ $\phi$ in whole region of the parameter space. Fig. \ref{dm_dd} (right panel) illustrates the relationship between the elastic scattering cross-section of DM with nucleons in a spin-independent manner as a function of the mass of DM. The red dashed and dashed-dotted lines in the plot represent the upper limits set by the XENON1T\cite{XENON:2017vdw,XENON:2018voc} and XENONnT\cite{XENON:2023sxq} collaborations, respectively. Also, the blue line represents the lower limit, which corresponds to the neutrino floor. In Fig. \ref{dm_lambda5}, we have illustrated the relationship between $\lambda_{5}$ and $M_{DM}$. The DM spans the entire spectrum of $\lambda_{5}$ values, ranging from -1 to $-10^{-5}$, as indicated by the green points.

\section{Leptogenesis in STSM}\label{lepto}
In this section, we delve into baryogenesis \textit{via} leptogenesis in STSM, considering two cases: one with two heavy fermions and the other with three, aimed at lowering the scale of leptogenesis.
\subsection{Two heavy fermion case}\label{lepto1}
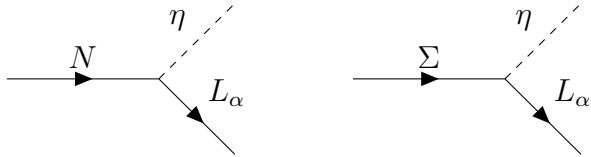
\begin{figure}[t]
    \centering
   \begin{tikzpicture}
\begin{feynman}
\vertex at (0,0) (i1);
\vertex at (-2,0) (i2);
\vertex at (1,1) (a);
\vertex at (1,-1) (b);
\diagram*{
(i2) -- [fermion, edge label=\(N\)] (i1), (i1) -- [scalar, edge label=\(\eta\)] (a), (i1) -- [fermion, edge label=\(L_{\alpha}\)] (b),
};
\end{feynman}
\end{tikzpicture}
\hspace{1cm}  \begin{tikzpicture}       
\begin{feynman}
\vertex at (0,0) (i1);
\vertex at (-2,0) (i2);
\vertex at (1,1) (a);
\vertex at (1,-1) (b);
\diagram*{
(i2) -- [fermion, edge label=\(\Sigma\)] (i1), (i1) -- [scalar, edge label=\(\eta\)] (a), (i1) -- [fermion, edge label=\(L_{\alpha}\)] (b),
};
\end{feynman}
\end{tikzpicture}
\caption{Tree-level Feynman diagrams for the decays $\left(N\rightarrow L_{\alpha}+\eta\right)$ and $\left(\Sigma\rightarrow L_{\alpha}+\eta\right)$.}
\label{fig1}
\end{figure}

\begin{figure}[t]
    \centering
\begin{tikzpicture}
\begin{feynman}
\vertex at (0,0) (i1);
\vertex at (-2,0) (i2);
\vertex at (2,1) (c);
\vertex at (2,-1) (d);
\vertex at (1,1) (a);
\vertex at (1,-1) (b);
\vertex at (1.5,-1.2) () {\(\eta\)};

\diagram*{
(i2) -- [fermion, edge label=\(N\)] (i1), (i1) -- [scalar, edge label=\(\eta\)] (a), (b) -- [fermion, edge label=\(L_{\beta}\)] (i1),
(a) --[plain, edge label=\(\Sigma\)] (b), (a) -- [fermion, edge label=\(L_{\alpha}\)] (c), (b) -- [scalar] (d),
};
\end{feynman}
\end{tikzpicture}\vspace{1cm}
\begin{tikzpicture}
\begin{feynman}
\vertex at (0,0) (i1);
\vertex at (-2,0) (i2);
\vertex at (2,1) (c);
\vertex at (2,-1) (d);
\vertex at (1,1) (a);
\vertex at (1,-1) (b);
\vertex at (1.5,-1.2) () {\(\eta\)};

\diagram*{
(i2) -- [fermion, edge label=\(\Sigma\)] (i1), (i1) -- [scalar, edge label=\(\eta\)] (a), (b) -- [fermion, edge label=\(L_{\beta}\)] (i1),
(a) --[plain, edge label=\(N\)] (b), (a) -- [fermion, edge label=\(L_{\alpha}\)] (c), (b) -- [scalar] (d),
};
\end{feynman}
\end{tikzpicture}
    \caption{Feynman diagrams illustrating the vertex corrections at the one-loop level in the decays of heavy fermions ($N$ and $\Sigma$).}
    \label{fig2}
\end{figure}
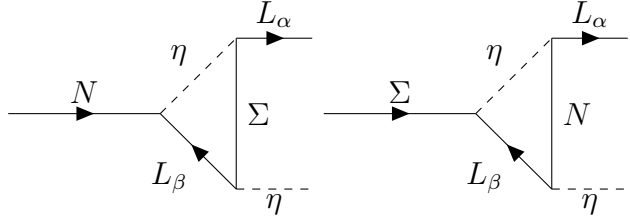

As previously mentioned, the present model allows for the emergence of a net lepton asymmetry through the out-of-equilibrium decay of both the $N$ and $\Sigma$.
The decay channels of singlet ($N$) and triplet ($\Sigma$) at tree and one-loop are shown in Figs. \ref{fig1} and \ref{fig2}. Here we consider the masses of $N$ and $\Sigma$ to be of similar order such that the decays of both contribute to leptogenesis. 
Much like the previously mentioned Davidson-Ibarra bound utilized in the context of Type-I seesaw leptogenesis, a similar lower limit can also be deduced in this scenario. The expressions for the $CP$ asymmetries resulting from the $N$ and $\Sigma$ decays, as shown in Figs. \ref{fig1} and \ref{fig2}, can be expressed as

\begin{eqnarray}\label{cp1}
\epsilon^N=\frac{1}{8 \pi\left(Y_N^{\dagger} Y_N\right)} \sum_{j \neq i} \operatorname{Im}\left[\left(Y_N^{\dagger} Y_\Sigma\right)_{i j}^2\right] \mathcal{F}\left(\frac{M_{\Sigma}^2}{M_{N}^2}\right),
\end{eqnarray}
and
\begin{eqnarray}\label{cp2}
\epsilon^{\Sigma}=\frac{1}{8 \pi\left(Y_{\Sigma}^{\dagger} Y_{\Sigma}\right)} \sum_{j \neq i} \operatorname{Im}\left[\left(Y_{\Sigma}^{\dagger} Y_N\right)_{i j}^2\right] \mathcal{F}\left(\frac{M_{N}^2}{M_{\Sigma}^2}\right),
\end{eqnarray}
where 
\begin{eqnarray}
Y_{N}=h_{\alpha 1}=\begin{pmatrix}
    Y_{N}^{1} \\
    Y_{N}^{2} \\
    Y_{N}^{3} \\
\end{pmatrix}, \hspace{1cm} Y_{\Sigma}=h_{\alpha 2}=\begin{pmatrix}
    Y_{\Sigma}^{1} \\
    Y_{\Sigma}^{2} \\
    Y_{\Sigma}^{3} \\
\end{pmatrix},
\end{eqnarray}
and
\begin{eqnarray}
\mathcal{F}(x)=\sqrt{x}\left[1+\frac{1}{1-x}+(1+x) \ln \left(\frac{x}{1+x}\right)\right] .
\end{eqnarray}

\noindent The relevant Boltzmann equations for leptogenesis can be written as
\begin{eqnarray}
    \dfrac{dn_{\Sigma}}{dz} & = & -D_{\Sigma} (n_{\Sigma}-n_{\Sigma}^{eq})-S_{A} (n_{\Sigma}^2-(n_{\Sigma}^{eq})^2),\label{bol1} \\
    \dfrac{dn_{N}}{dz} & = & -D_{N} (n_{N}-n_{N}^{eq}),\label{bol2} \\\nonumber
    \dfrac{dn_{B-L}}{dz} & = & -\epsilon^{\Sigma} D_{\Sigma} (n_{\Sigma}-n_{\Sigma}^{eq})-\epsilon^{N}D_{N} (n_{N}-n_{N}^{eq})-W_{\rm ID}^{\Sigma}n_{B-L}-W_{\rm ID}^{N}n_{B-L}\\
    & & - W^{}_{\Delta L} n_{B-L}.\label{bol3}
\end{eqnarray}
\noindent Here, $n_{N}^{eq}$ and $n_{\Sigma}^{eq}$ are the equilibrium number density of $N$ and $\Sigma$ respectively. Also, $n_{N}$, $n_{\Sigma}$ and $n_{B-L}$ are the comoving number densities of $N$, $\Sigma$ and $B-L$ asymmetry respectively. $D_{N}$ and $D_{\Sigma}$ are the decay terms for $N$ and $\Sigma$ decays which are given by
\begin{eqnarray}
    D_{N} & = & K_{N}z\dfrac{\kappa_{1}(z)}{\kappa_{2}(z)}, \\
  D_{\Sigma} & = & K_{\Sigma} \bigg( \dfrac{M_{\Sigma}}{M_{N}}  z\bigg) \dfrac{\kappa_{1} \bigg(  \dfrac{M_{\Sigma}}{M_{N}}z \bigg)}{\kappa_{2} \bigg(  \dfrac{M_{\Sigma}}{M_{N}}z \bigg)},
\end{eqnarray}
where $z=\frac{M_{N}}{T}$ with $T$ is temperature of thermal bath and
\begin{eqnarray}
  K_{N} & = & \dfrac{\Gamma_{N}}{\textbf{H}(T=M_{N})}, \\\
  K_{\Sigma} & = & \dfrac{\Gamma_{\Sigma}}{\textbf{H}(T=M_{\Sigma})},
\end{eqnarray}
are the decay parameters for the $N$ and $\Sigma$ decay, respectively. $\Gamma_{N,\Sigma}$ are the decay widths given by
\begin{eqnarray}
  \Gamma_{N} & = & \dfrac{M_{N}}{8\pi} (Y_{N}^{\dagger}Y_{N}) \bigg(  1-\dfrac{m_{\eta}^{2}}{M_{N}^{2}}\bigg), \\
  \Gamma_{\Sigma} & = & \dfrac{M_{\Sigma}}{8 \pi} (Y_{\Sigma}^{\dagger}Y_{\Sigma}) \bigg(  1- \dfrac{m_{\eta}^{2}}{M_{\Sigma}^{2}}  \bigg).
\end{eqnarray}
\noindent In the early Universe, dominated by radiation, the Hubble parameter can be expressed in terms of temperature as $H=2\sqrt{\frac{\pi^3 g_{*}}{45}}\frac{T^2}{M_{pl}}$, with $g_{*}=124$ is the effective number of relativistic degrees of freedom and $M_{Pl}$ is the Planck mass given by $M_{Pl}=1.22 \times 10^{19}$ GeV. The washout term arises from the combined effect of inverse decays and $L$ violating washout processes. The total washout term is given by  $W^{}_{\textit{\rm Total}}=W^{N,\Sigma}_{\textit{\rm ID}}+W_{\rm \Delta L}$, where $W_{\rm ID}^{N}$ and $W_{\rm ID}^{\Sigma}$ are the inverse decay terms for $N$ and $\Sigma$ decay, respectively and are given by 
\begin{eqnarray}
    W_{\rm ID}^{N} & = & \dfrac{1}{4}K_{N}z^{3}\kappa_{1}(z), \\
    W_{\rm ID}^{\Sigma} & = & \dfrac{1}{4}K_{\Sigma} \bigg( \dfrac{M_{\Sigma}}{M_{N}}z \bigg)^{3}\kappa_{1} \bigg( \dfrac{M_{\Sigma}}{M_{N}}z  \bigg). 
\end{eqnarray}

\noindent Apart from the inverse decays, the other contribution to washout $W_{\rm \Delta L}$ originates from scatterings that violate the lepton number. The scattering washouts due to $l\eta \longrightarrow \bar{l} \eta^{*}$, $ll \longrightarrow \eta^{*} \eta^{*}$ process is given by
\begin{eqnarray}
&&W_{\rm \Delta L}=\frac{0.585\times M_{pl}}{g_{l}\times g_{*}z^2 v^4}\left(\frac{2 \pi^2}{\lambda_{5}}\right)^2 M_{N}\overline{m}_{\varkappa_{N}}^2+\frac{0.585\times M_{pl}}{g_{l}\times g_{*}z^2 v^4}\left(\frac{2 \pi^2}{\lambda_{5}}\right)^2 M_{\Sigma}\overline{m}_{\varkappa_{\Sigma}}^2.
\end{eqnarray}
Here, $g_{l}$ represents the intrinsic degrees of freedom of SM leptons, while $\overline{m}_\varkappa$ is defined as an effective neutrino mass parameter and in the limit of $m_{1}$=0, given by
\begin{eqnarray}
\overline{m}_{\varkappa_{N,\Sigma}}^2=\varkappa_{2}^2 m_{2}^{2}+\varkappa_{3}^2 m_{3}^{2}
\end{eqnarray}
with $m_{i}^s$ being the light neutrino mass eigenvalues and $\varkappa$ is defined as:
\begin{eqnarray}
\varkappa_{2,3}=\frac{M_{N,\Sigma}^{2}}{8\left(m^{2}_{\eta^{R}}-m^{2}_{\eta^{I}}\right)}[L_{i}(m_{\eta^R}^2)-L_{i}(m_{\eta^I}^2)].
\end{eqnarray}

\noindent The gauge boson mediated scattering term, $S_{A}$, for $\Sigma$ and can be identified as
\begin{eqnarray}
S_A=\left(\frac{0.032 \sqrt{g_{*}}M_{pl}}{g^2_{\Sigma}M_{\Sigma}}\right)\left(\frac{I(z)}{z\times \kappa_{2} \bigg(  \dfrac{M_{\Sigma}}{M_{N}}z \bigg)^2}\right),
\end{eqnarray}
where
\begin{eqnarray}
I(z)=\int_{4}^{\infty} \sqrt{x} \kappa_{1}\bigg(z\sqrt{x}\bigg)\hat{\sigma}_{A}(x) \,dx\  .
\end{eqnarray}
Here, the $\hat{\sigma}(x)$ is cross-section for the
gauge boson mediated processes given by\cite{Biswas:2023azl}
\begin{eqnarray}
\hat{\sigma}_A=\frac{6 g^4}{72 \pi}\left[\frac{45}{2} r(x)-\frac{27}{2} r(x)^3-\left\{9\left(r(x)^2-2\right)+18\left(r(x)^2-1\right)^2\right\} \ln \left(\frac{1+r(x)}{1-r(x)}\right)\right],
\label{41}
\end{eqnarray}
where $r(x)=\sqrt{1-4/x}$. Eqn. (\ref{41}) includes $\Sigma$, $\Sigma$ → all possible fermion doublets, Higgs, gauge bosons. After solving the Boltzmann equations mentioned in Eqns. (\ref{bol1}), (\ref{bol2}), and (\ref{bol3}), the relation
\begin{eqnarray}
\eta_{B}=\frac{3 g_{*}^0}{4 g_*} a_{sphl} n_{B-L}\simeq 9.2\times 10^{-3} n_{B-L},
\end{eqnarray}
converts $B-L$ asymmetry, $n_{B-L}$, into the observed baryon-to-photon ratio, prior to electroweak sphaleron freeze-out. Here $a_{sphl}=\frac{8}{23}$ is the sphaleron conversion factor, and $g_{*}^{0}$ represents the effective count of relativistic degrees of freedom during the epoch of recombination.

\subsection{Three heavy fermions case}\label{lepto2}
Here we consider two singlet heavy fermions ($N_{1}$ and $N_{2}$) with one triplet fermion $\Sigma$. $M_{N_{1}}$, $M_{N_{2}}$ and $M_{\Sigma}$ are the masses of $N_{1}$, $N_{2}$ and $\Sigma$, respectively. Due to the presence of an additional singlet fermion, the Yukawa matrix $h$ takes the form

\begin{eqnarray}
    h & = & \begin{pmatrix}
        Y_{N_{1}}^{1} & Y_{\Sigma}^{1} & Y_{N_{2}}^{1}\\ Y_{N_{1}}^{2} & Y_{\Sigma}^{2} &  Y_{N_{2}}^{2} \\
        Y_{N_{1}}^{3} & Y_{\Sigma}^{3} & Y_{N_{2}}^{3} 
    \end{pmatrix} V^{T}(\alpha) .
    \label{yukawa2}
\end{eqnarray}

\noindent Here, we perform the analysis assuming the hierarchy $M_{N_{1}} \lesssim M_{\Sigma} \ll M_{N_{2}}$. Since $N_{2}$ is assumed to be much heavier compared to the other two fermions, the lepton asymmetry generated from the decay of $N_{2}$ is washed-out by the inverse decays of $N_{1}$ and $\Sigma$. Effectively, only $N_{1}$ and $\Sigma$ will generate the lepton asymmetry. The relevant CP asymmetry parameter due to the decay of $N_{1}$ and $\Sigma$ are found to be
\begin{eqnarray}
    \epsilon_{N_{1}} & = & \frac{1}{8 \pi\left(Y_{N_{1}}^{\dagger} Y_{N_{1}}\right)} \left[ \operatorname{Im}\left[\left(Y_{N_{1}}^{\dagger} Y_\Sigma\right)_{}^2\right] \mathcal{F}\left(\frac{M_{\Sigma}^2}{M_{N_{1}}^2}\right) + \left(Y_{N_{1}}^{\dagger} Y_\Sigma\right)_{}^2 \mathcal{F}\left(\frac{M_{\Sigma}^2}{M_{N_{1}}^2}\right) \right], \\
     \epsilon_{\Sigma} & = & \frac{1}{8 \pi\left(Y_{\Sigma}^{\dagger} Y_{\Sigma}\right)} \left[ \operatorname{Im}\left[\left(Y_{\Sigma}^{\dagger} Y_{N_{1}}\right)_{}^2\right] \mathcal{F}\left(\frac{M_{N_{1}}^2}{M_{\Sigma}^2}\right) + \left(Y_{\Sigma}^{\dagger} Y_{N_{2}}\right)_{}^2 \mathcal{F}\left(\frac{M_{N_{2}}^2}{M_{\Sigma}^2}\right) \right].
\end{eqnarray}

\noindent The relevant Boltzmann equations in this scenario are 
\begin{eqnarray}
\dfrac{dn_{N_{1}}}{dz} &= & -D_{N} \left(n_{N_{1}}-n^{eq}_{N_{1}}\right),\label{bol21}\\
\dfrac{dn_{\Sigma}}{dz} & = & -D_{\Sigma} (n_{\Sigma}-n_{\Sigma}^{eq})-S_{A}(n_{\Sigma}^{2}-(n_{\Sigma}^{eq})^{2}),\label{bol22}\\
\dfrac{dn_{B-L}}{dz} & = & -\epsilon_{N_{1}}D_{N_{1}}(n_{N_{1}}-n_{N_{1}}^{eq})-\epsilon_{\Sigma}D_{\Sigma}(n_{\Sigma}-n_{\Sigma}^{eq})-W_{\rm ID}^{N_{1}}n_{B-L}-W_{\rm ID}^{\Sigma} n_{B-L} \nonumber \\ & & - W^{}_{\Delta L} n_{B-L}.\label{bol23}
\end{eqnarray}

\noindent Here, the decay terms for $N_{1}$ and $\Sigma$ are given by
\begin{eqnarray}
    D_{N_{1}} & = & K_{N_{1}}\dfrac{\kappa_{1}(z)}{\kappa_{2}(z)}, \\
    D_{\Sigma} & = & K_{\Sigma} \left( \dfrac{M_{\Sigma}}{M_{N_{1}}}z \right) \dfrac{\kappa_{1}\left( \dfrac{M_{\Sigma}}{M_{N_{1}}}z\right)}{k_{2} \left( \dfrac{M_{\Sigma}}{M_{N_{1}}}z \right)}.
\end{eqnarray}
and inverse decay terms for $N_{1}$ and $\Sigma$ are defined as
\begin{eqnarray}
    W_{\rm ID}^{N_{1}} & = & \dfrac{1}{4}K_{N_{1}}z^{3}\kappa_{1}(z), \\
    W_{ID}^{\Sigma} & = & \dfrac{1}{4} K_{\Sigma} \left( \dfrac{M_{\Sigma}}{M_{N_{1}}}z \right)^{3} \kappa_{1} \left( \dfrac{M_{\Sigma}}{M_{N_{1}}}z \right). 
\end{eqnarray}

\begin{figure}[t]
    \centering
    \includegraphics[scale=0.3]{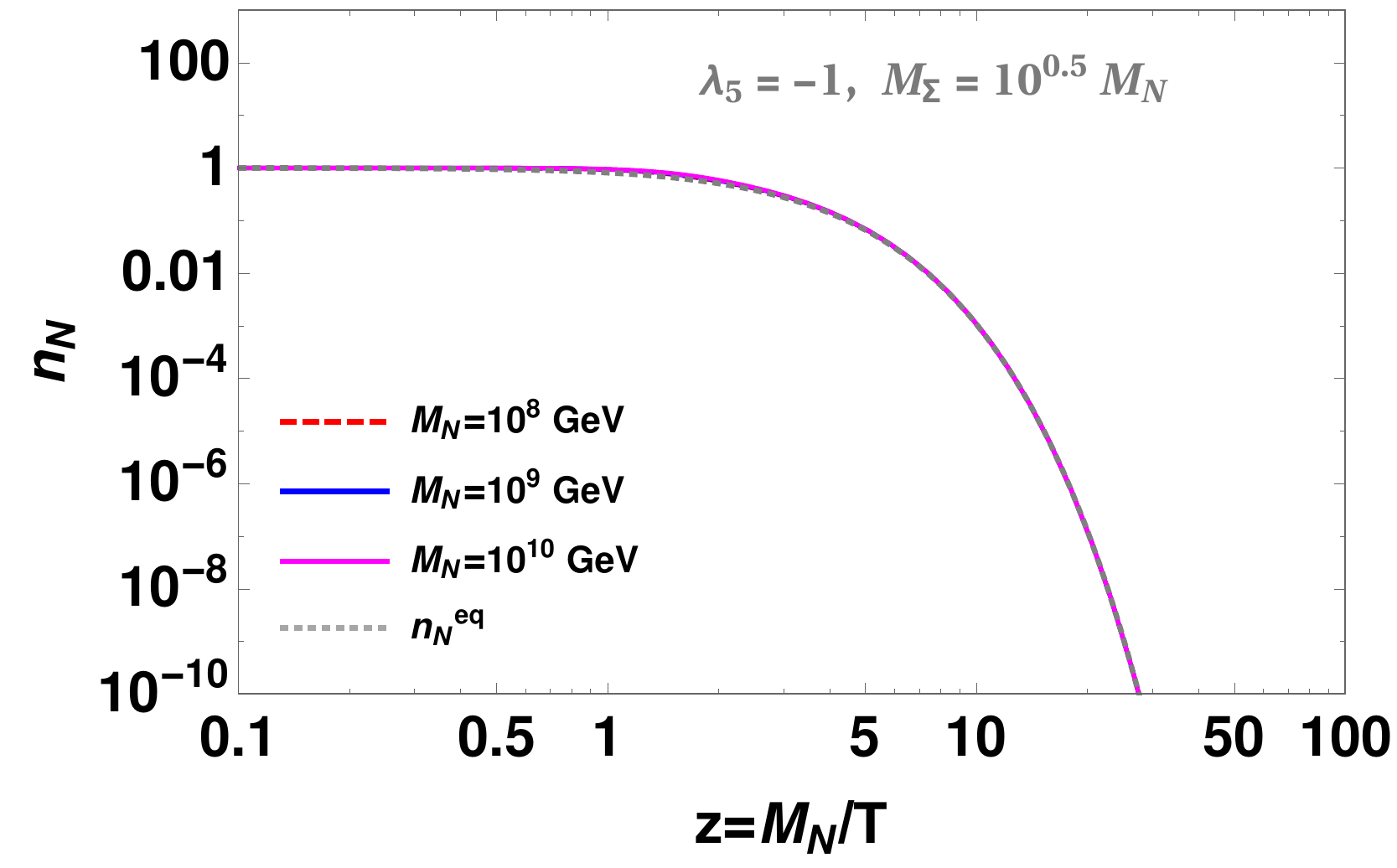}
\includegraphics[scale=0.3]{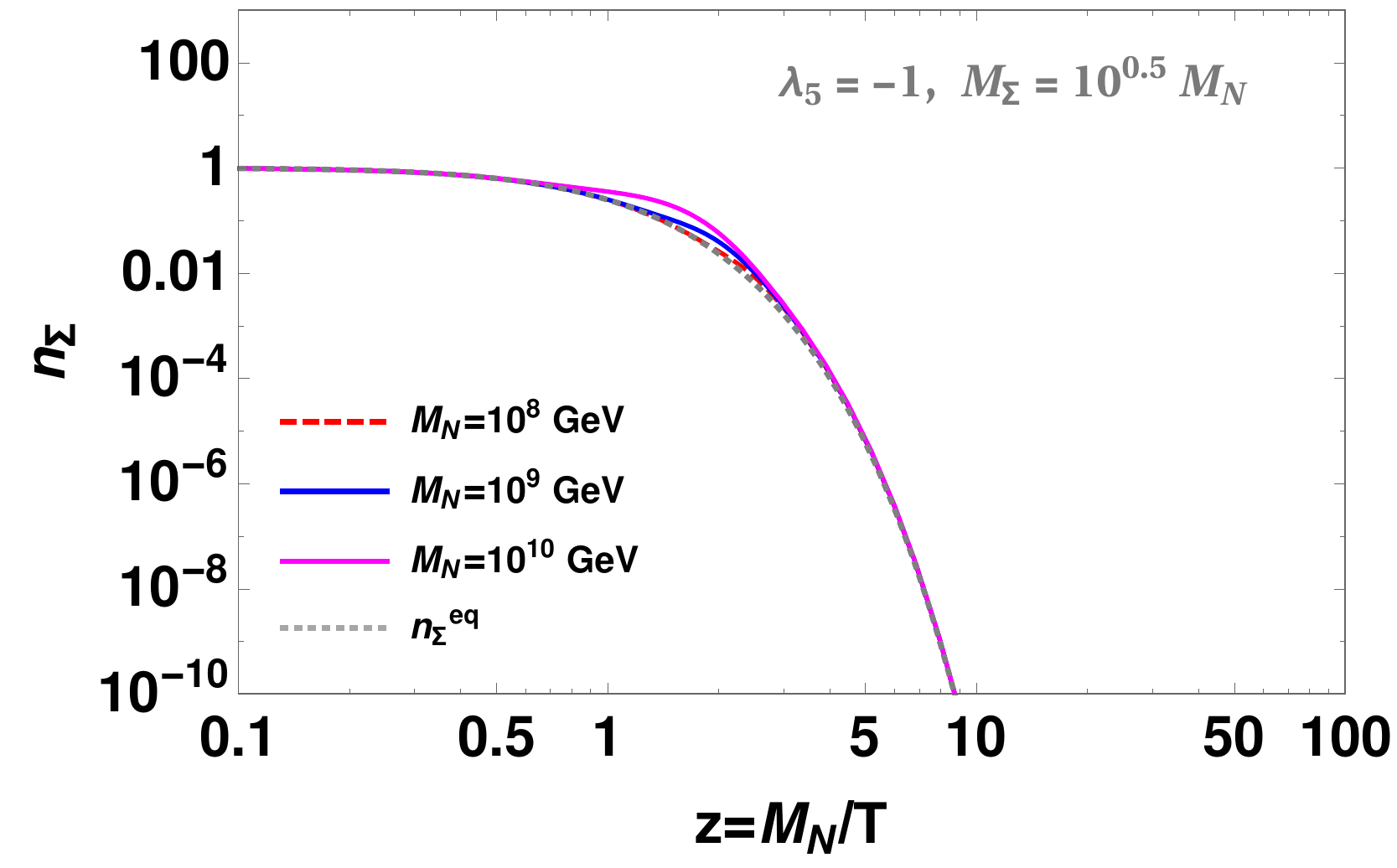}\\
    \includegraphics[scale=0.3]{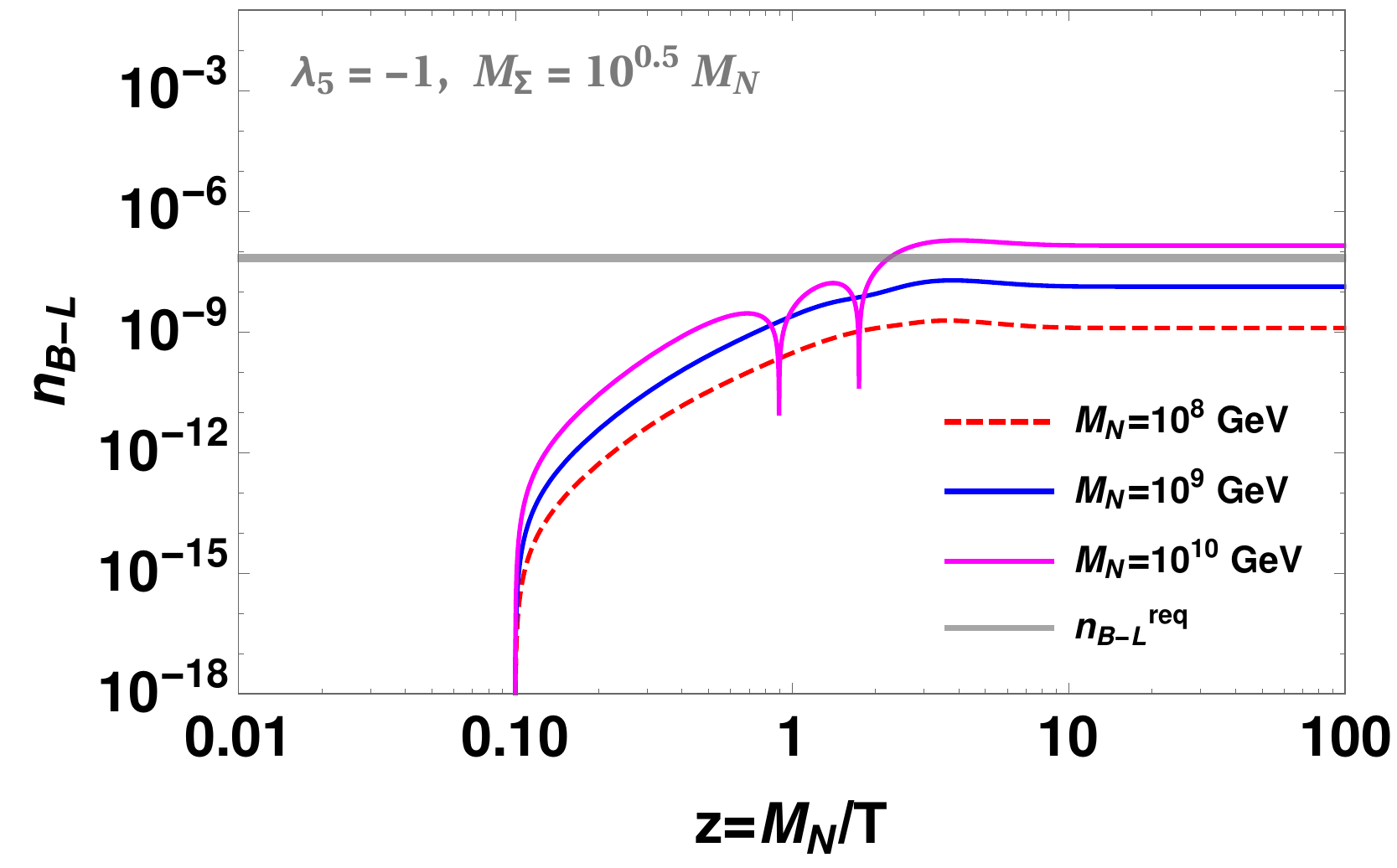}
    \caption{Comoving number density of $N$ (upper left panel), $\Sigma$ (upper right panel) and $B-L$ (lower panel) with different values of $M_{N}$ in two moderately hierarchical heavy fermions case. In the lower panel, the grey horizontal line represents the experimentally required value for the $B-L$ asymmetry. }
    \label{bl1}
\end{figure}

\begin{figure}[h]
    \centering
    \includegraphics[scale=0.3]{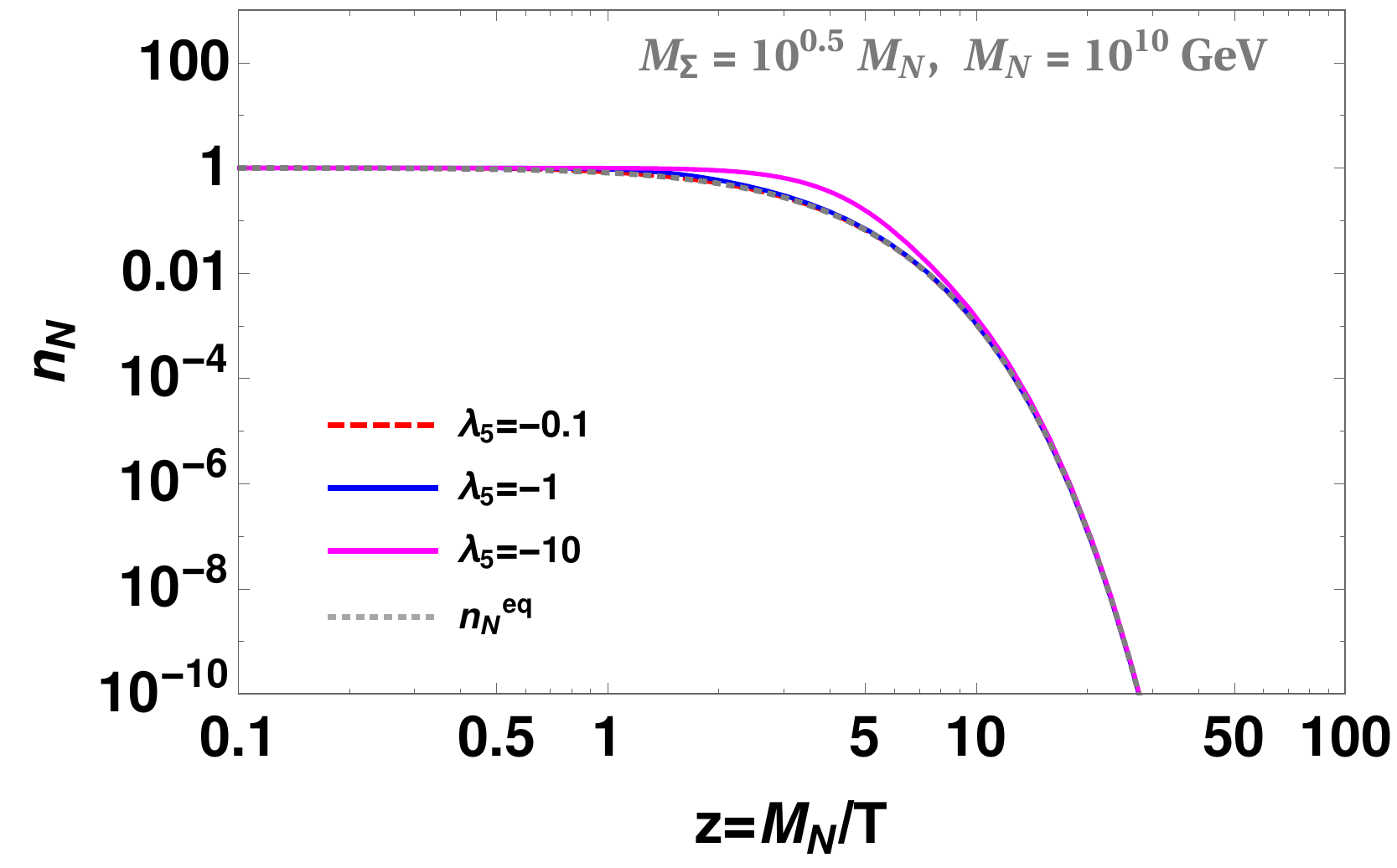}
\includegraphics[scale=0.3]{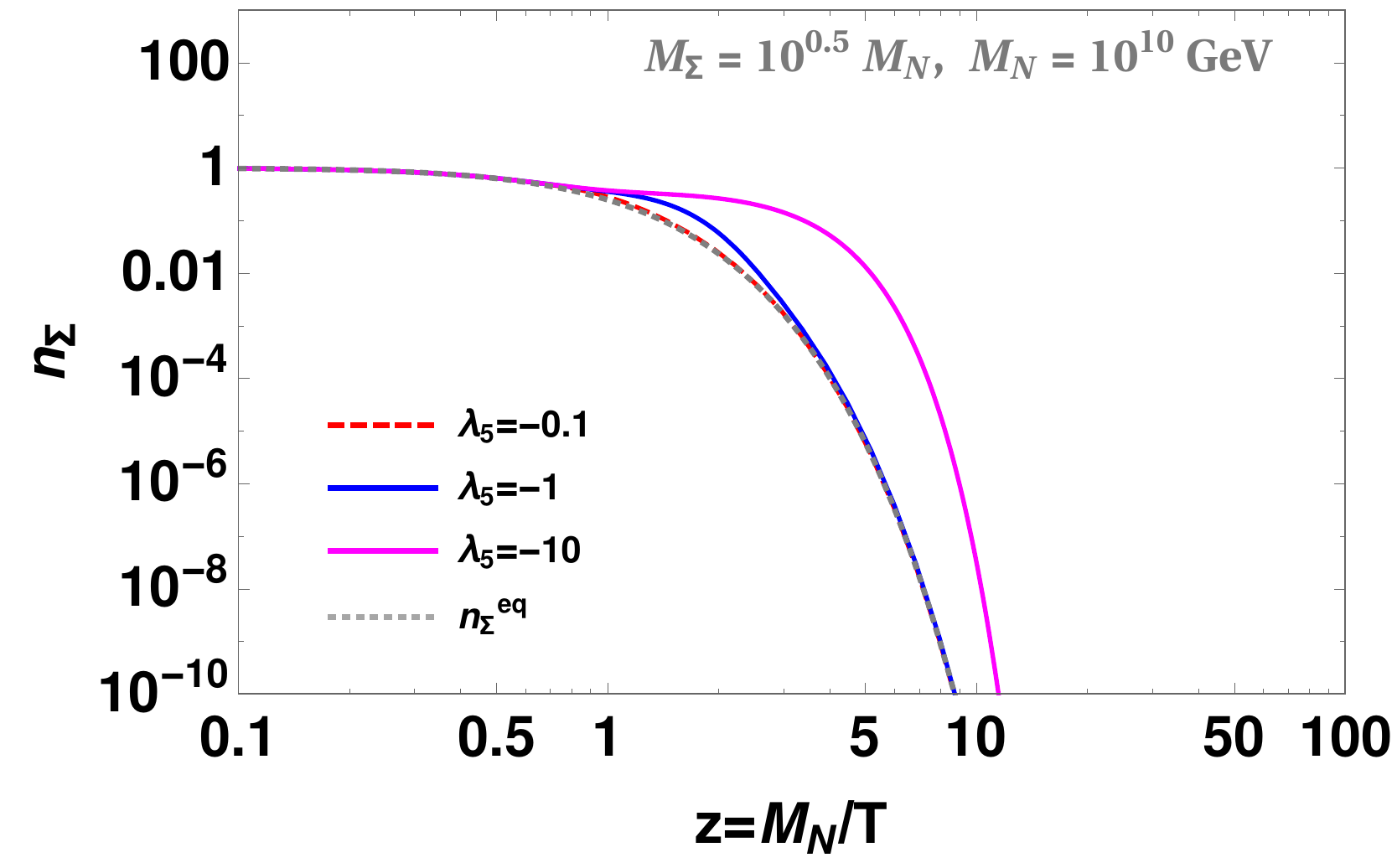}\\
    \includegraphics[scale=0.3]{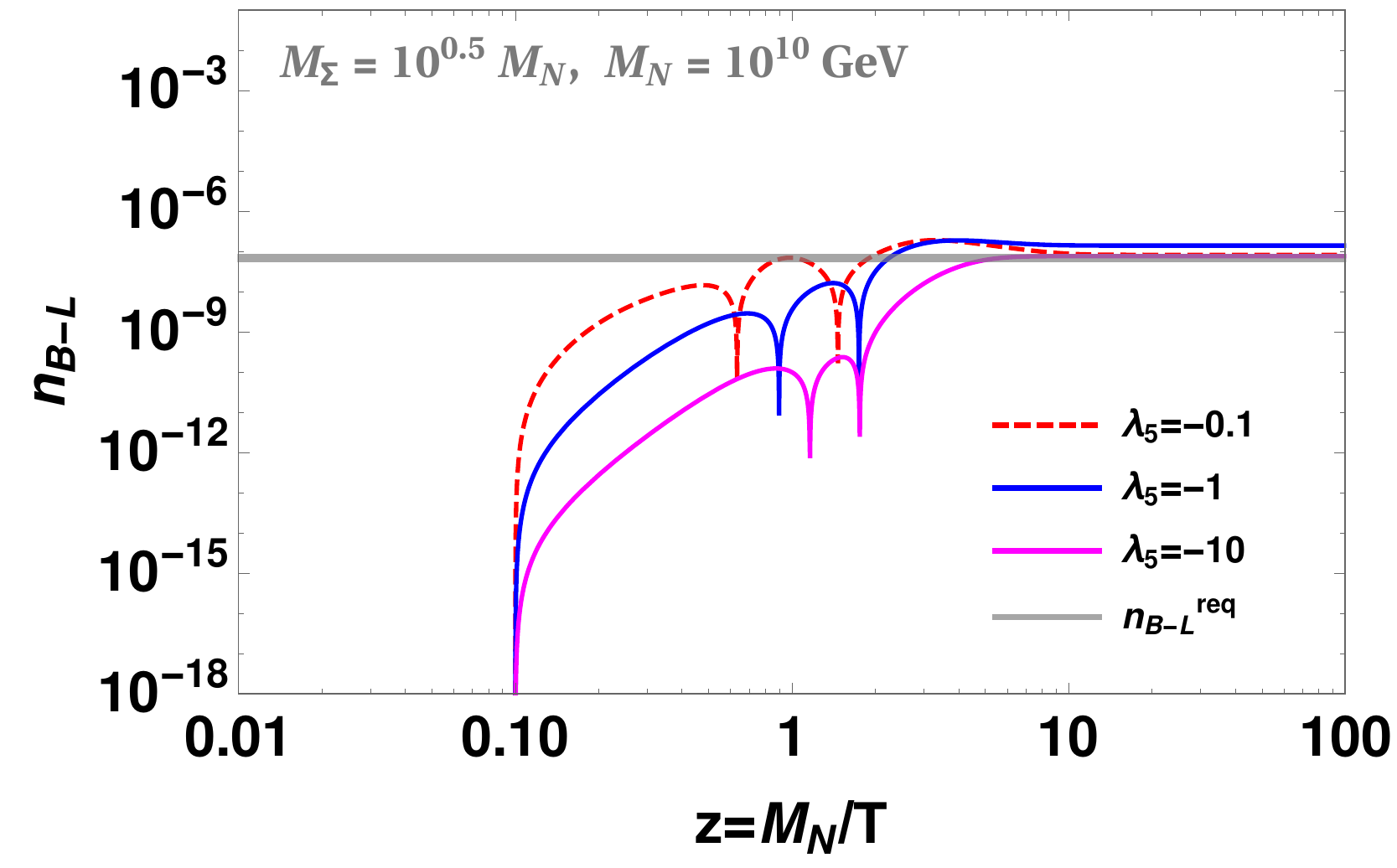}
    \caption{Comoving number density of $N$ (upper left panel), $\Sigma$ (upper right panel) and $B-L$ (lower panel) with different values of $\lambda_{5}$ in two moderately hierarchical heavy fermions case. In the lower panel, the grey horizontal line represents the experimentally required value for the $B-L$ asymmetry.}
    \label{bl2}
\end{figure}

\begin{figure}[t]
    \centering
    \includegraphics[scale=0.3]{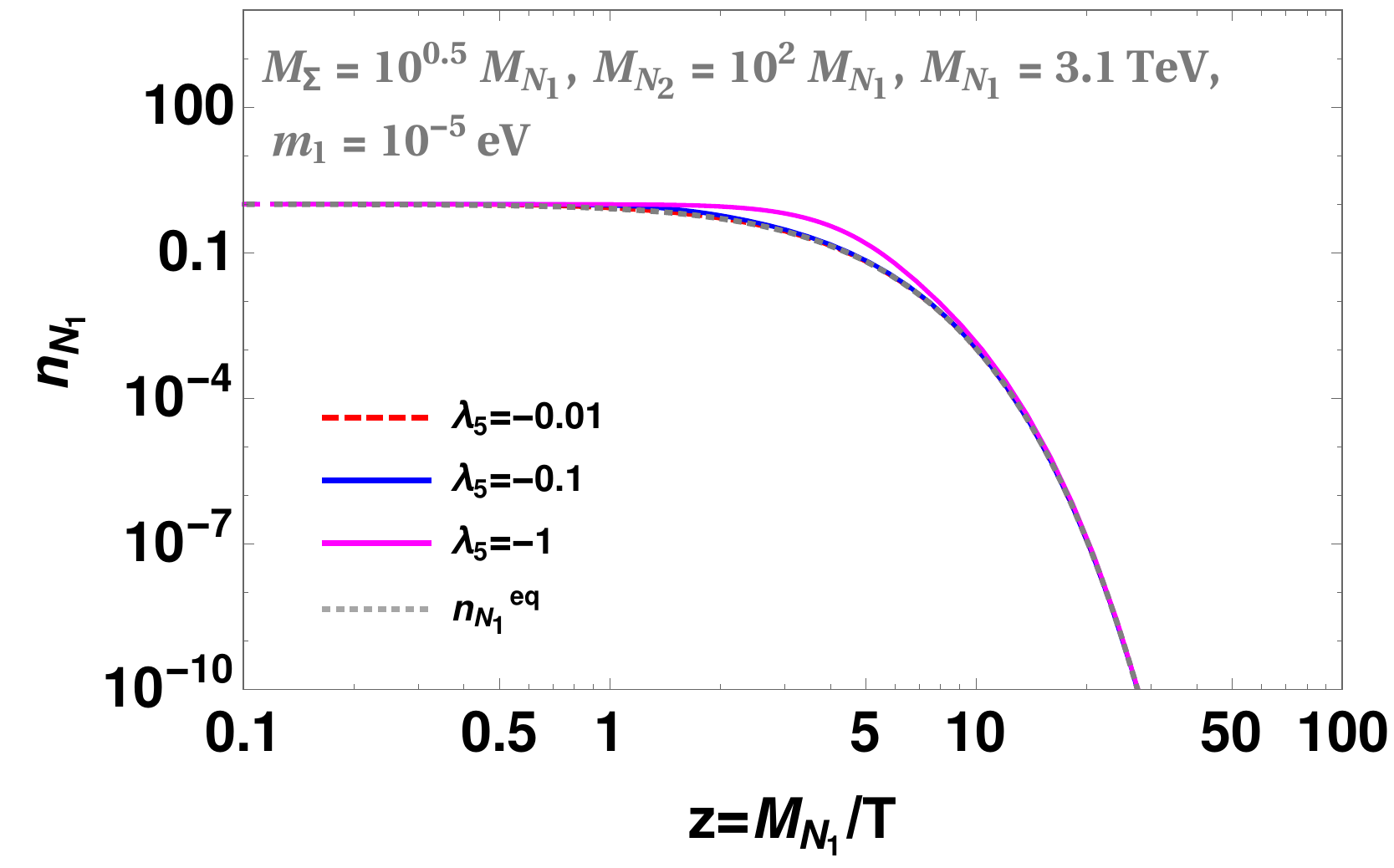}
    \includegraphics[scale=0.3]{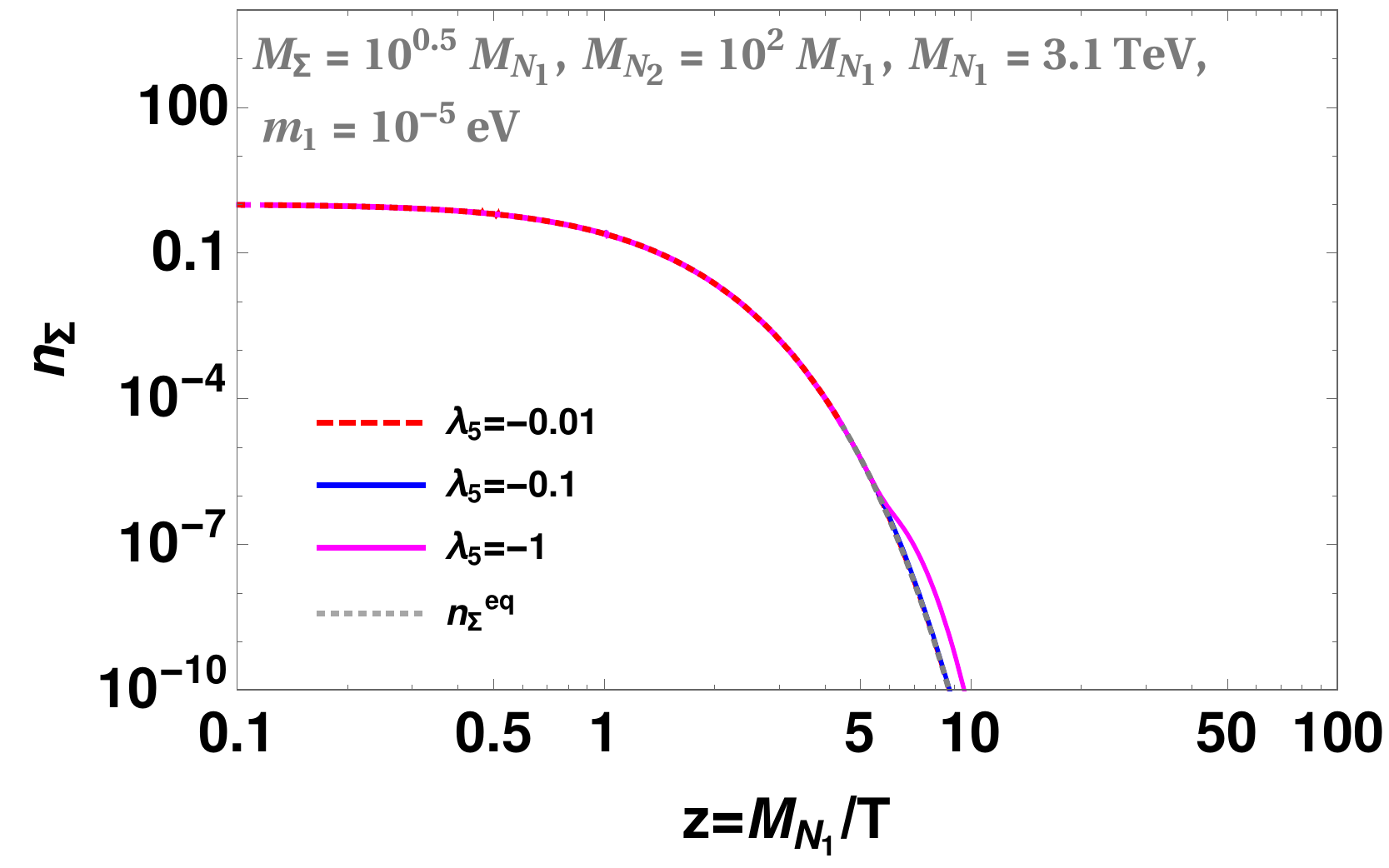}\\
    \includegraphics[scale=0.3]{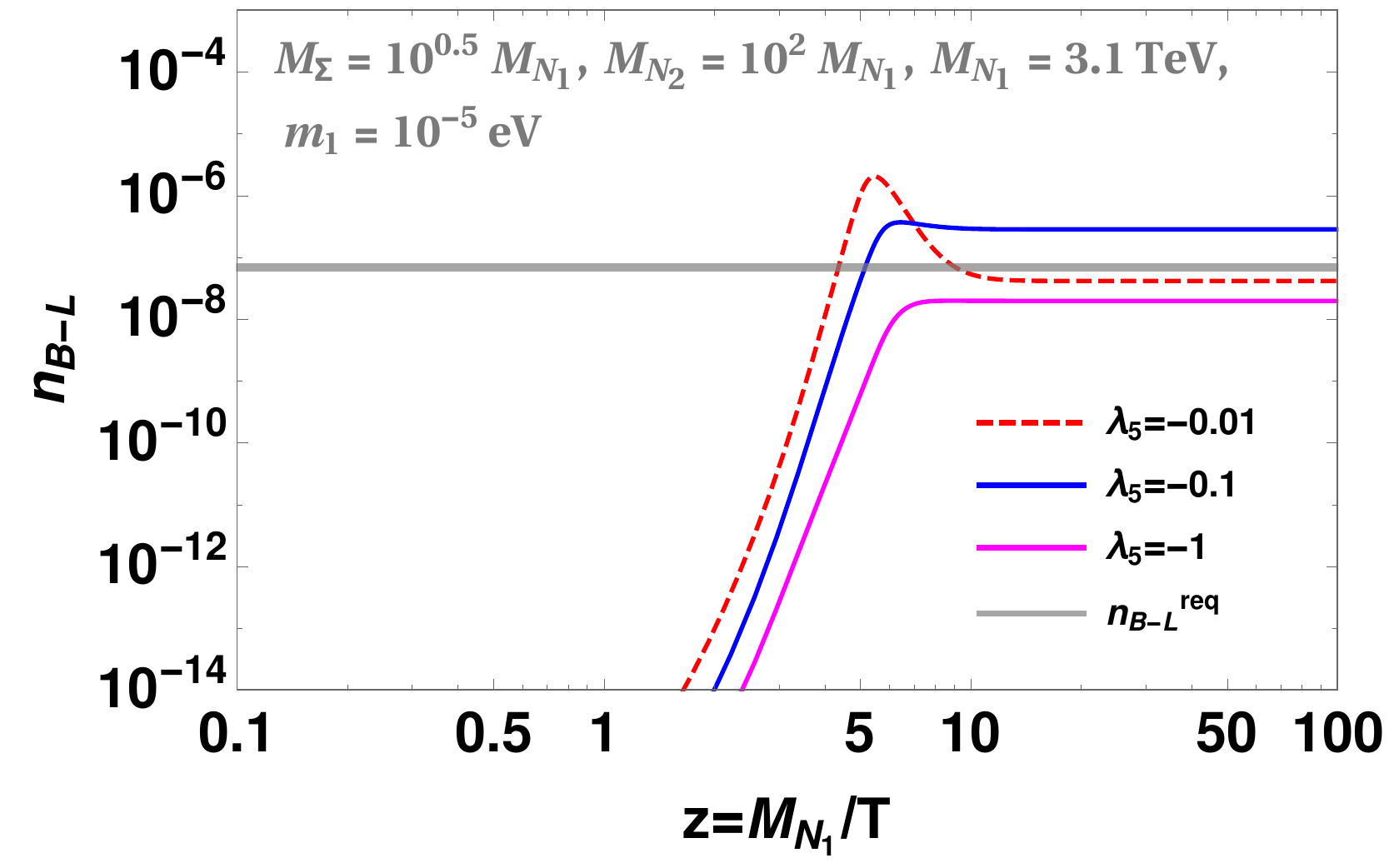}
    \caption{Comoving Number density of $N_{1}$ (top left panel plot), $\Sigma$ (top right panel plot) and $B-L$ (lower panel plot) with $z=M_{N_{1}}/T$ for different values of $\lambda_{5}$ in three heavy fermions case. In the lower panel, the grey horizontal line represents the experimentally required value for the $B-L$ asymmetry.}
    \label{bl3}
\end{figure}

\begin{figure}[t]
    \centering
    \includegraphics[scale=0.3]{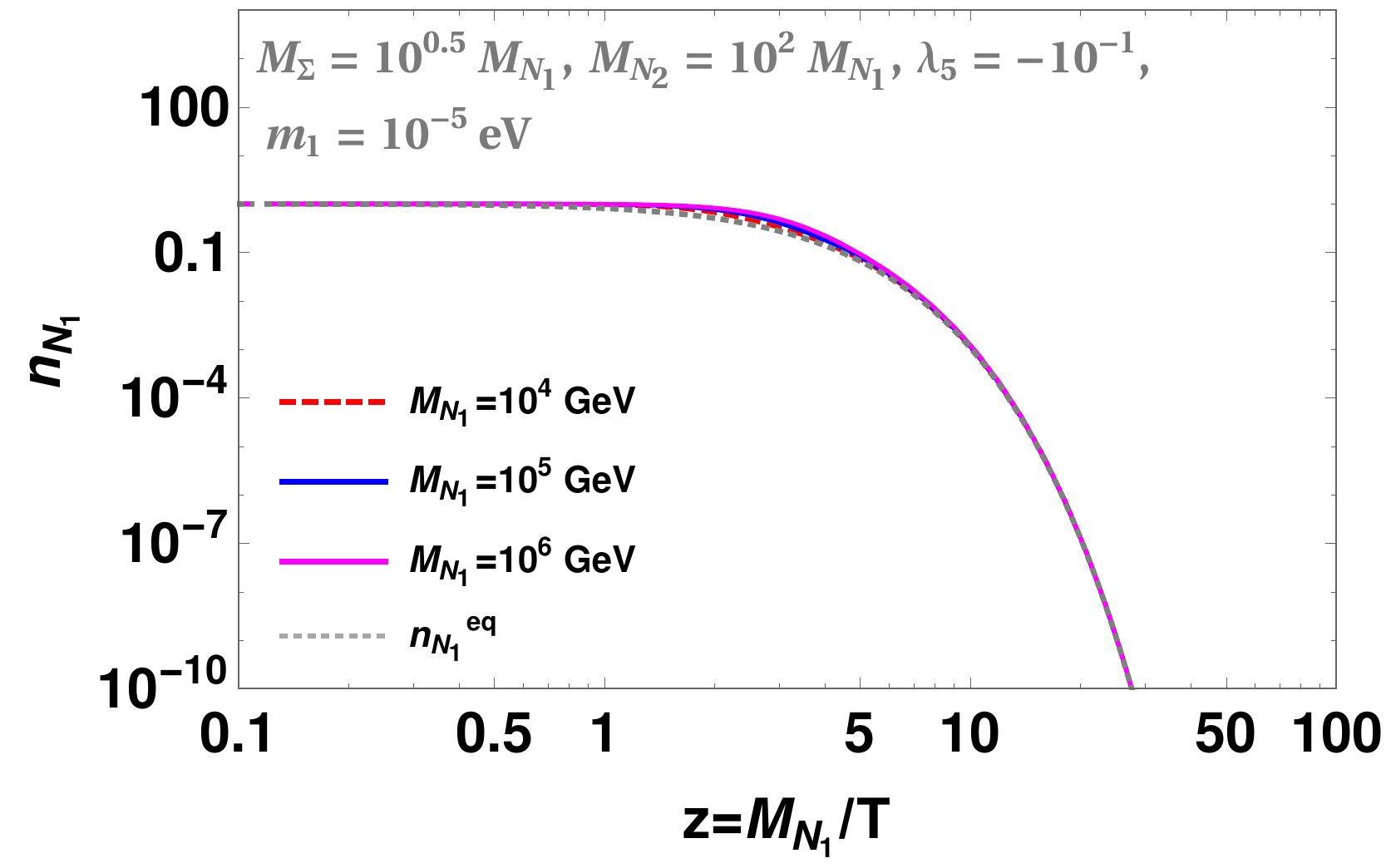}
    \includegraphics[scale=0.3]{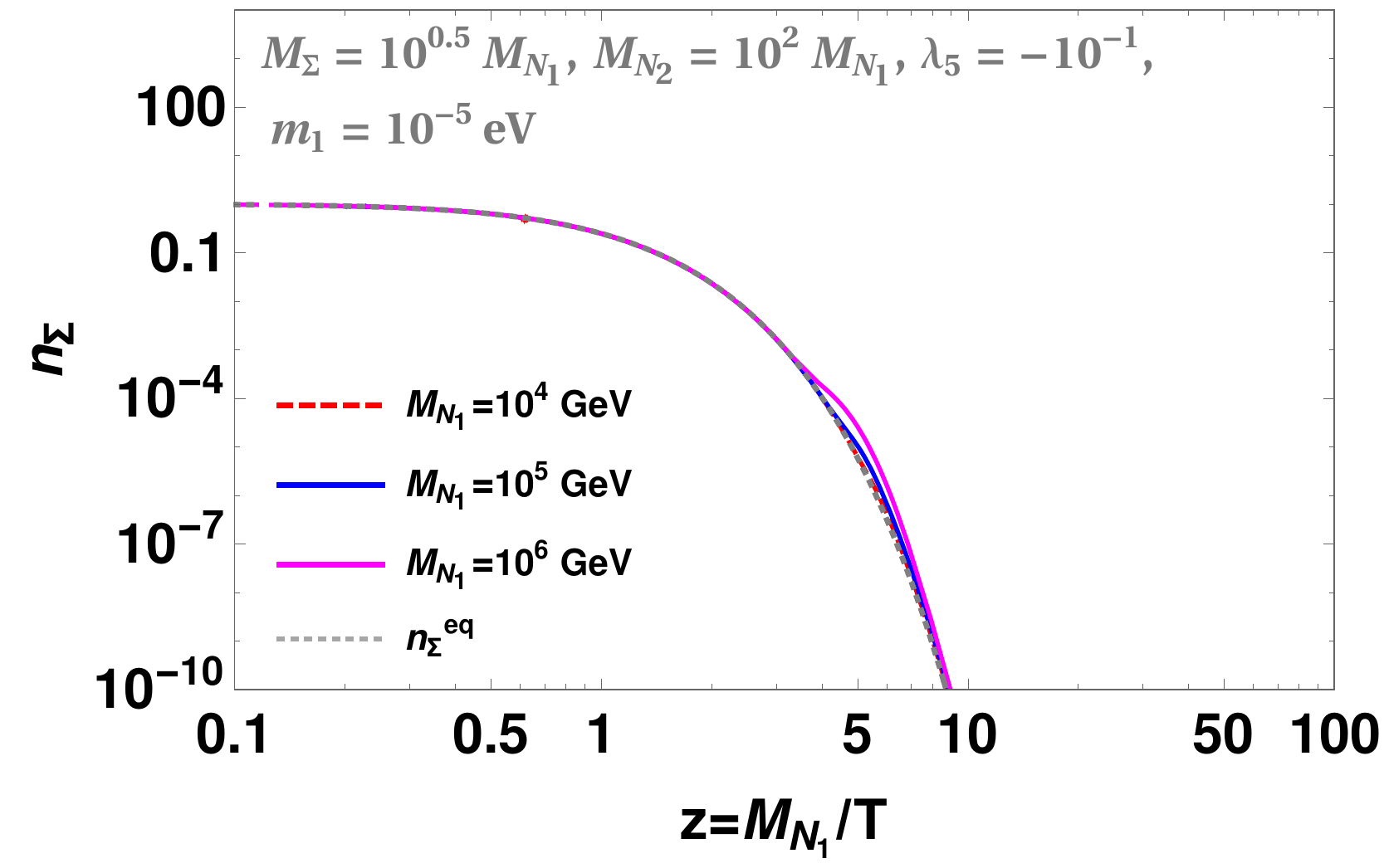}\\
    \includegraphics[scale=0.3]{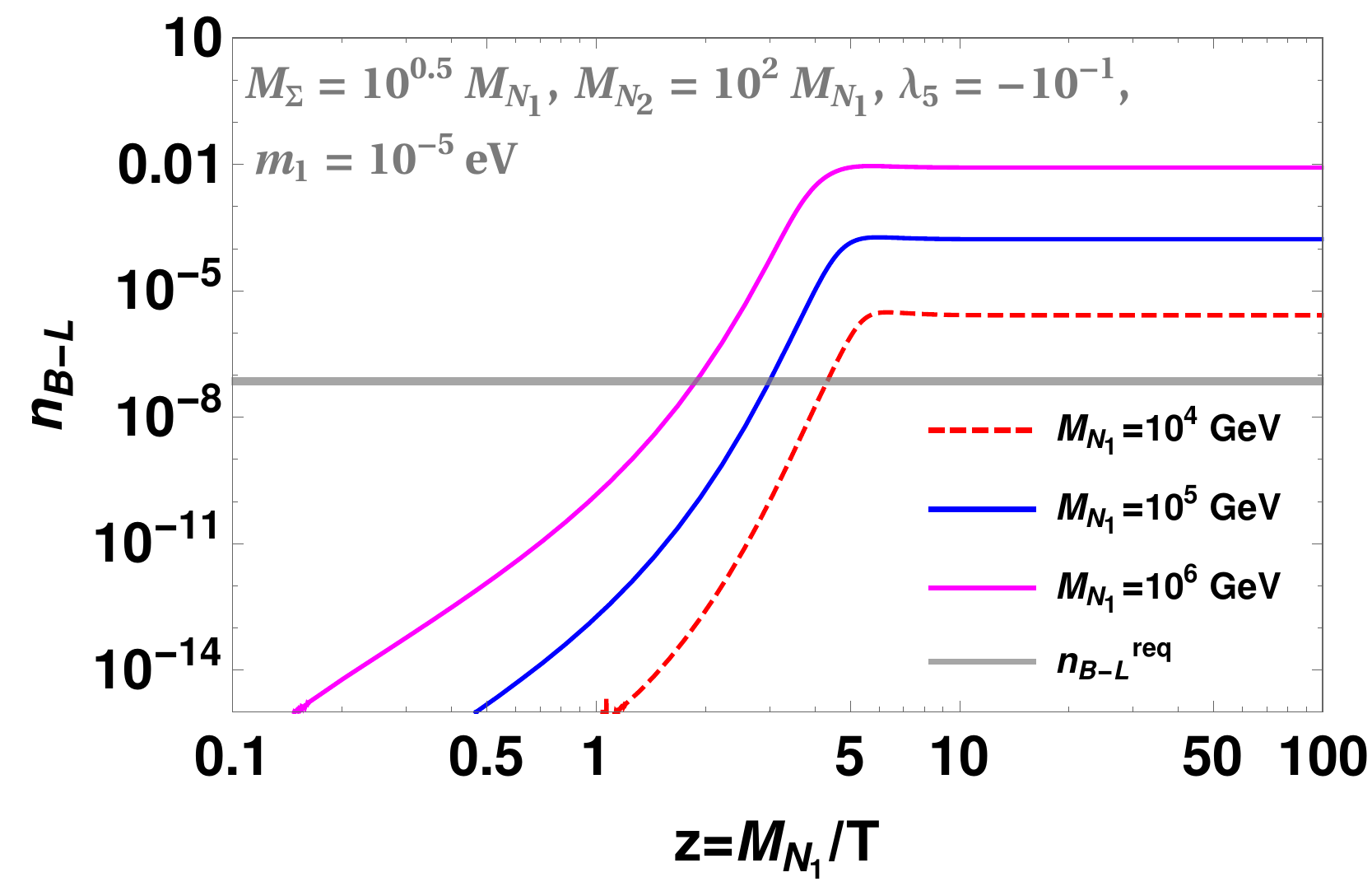}
    \caption{Comoving Number density of $N_{1}$ (upper left panel), $\Sigma$ (upper right panel) and $B-L$ (lower panel) with $z=M_{N}/T$ for different values of $M_{N_{1}}$ in three heavy fermions case. In the lower panel, the grey horizontal line represents the experimentally required value for the $B-L$ asymmetry.}
    \label{bl4}
\end{figure}

\begin{figure}[t]
    \centering
    \includegraphics[width=11cm,height=7cm]{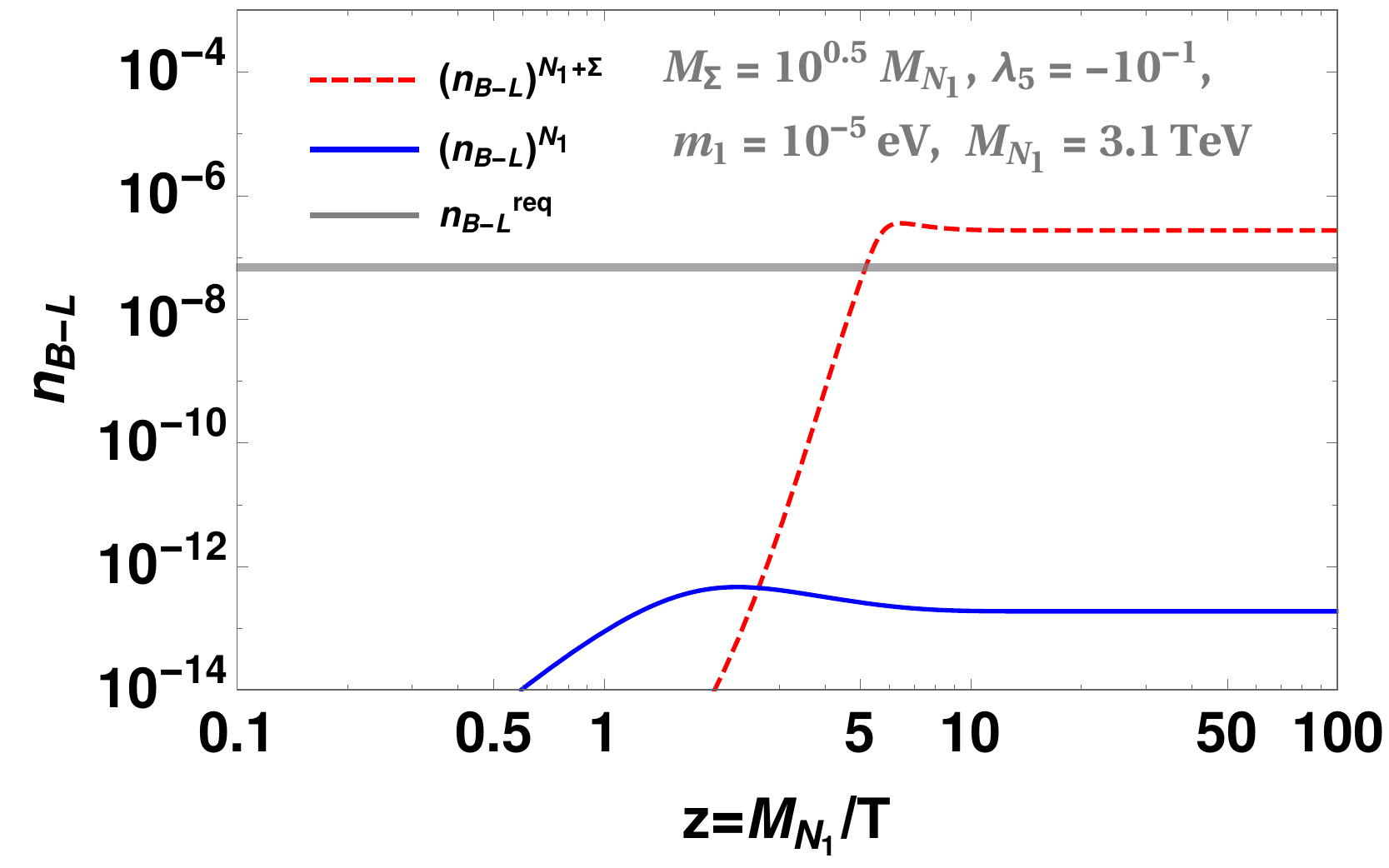}
    \caption{Evolution of $B-L$ asymmetry for $N_{1}+\Sigma$ (red dashed line) and $N_{1}$ (blue line) as a function of $z=M_{N_{1}}/T$. The grey horizontal line represents the experimentally required value for the $B-L$ asymmetry. }
    \label{compare}
\end{figure}

\subsection{Numerical Analysis and Discussion}\label{result}
We study the leptogenesis in the STSM where we first consider the minimal scenario with two moderately hierarchical ($M_{N}\lesssim M_{\Sigma}$) heavy fermions, $N$ and $\Sigma$.  This results in the lightest neutrino remaining massless. In order to satisfy Sakharov's third condition, the lepton asymmetry must be generated by the out-of-equilibrium decay of heavy fermions. This asymmetry is then converted into the baryon asymmetry $via$ ($B+L$) violating sphaleron processes. As the Universe evolves, characterized by a decrease in temperature or increase in $z$, $N$ and $\Sigma$ begin decaying into SM particles reducing their respective comoving number densities ($n_{N}$ and $n_{\Sigma}$). Consequently, this process occurs concomitantly with an escalation in lepton asymmetry $n_{B-L}$, as depicted in Figs. \ref{bl1} and \ref{bl2}. At fixed $\lambda_{5}=-1$ and $M_{\Sigma}=10^{0.5}M_{N}$, $n_N$ and $n_{\Sigma}$ are very close to $n^{eq}_{N}$ and $n^{eq}_{\Sigma}$ as shown in the upper panel of Fig. \ref{bl1}. Due to the presence of only two heavy fermions generating masses of the neutrinos, the Yukawa couplings for the heavy fermions are large enough such that the decay and inverse decays keep their abundances very close to equilibrium abundance. This is explained in detail in Appendix \ref{appendix1}. Due to the same reason, we are necessarily in a strong washout region ($K_{N,\Sigma} >> 1$) with two heavy fermions. This leads to a high-scale leptogenesis scenario where $M_{N,\Sigma} \geq 10^{9}$ GeV. In Fig. \ref{bl1}, one can see that $n_{\Sigma}$ deviates more from its equilibrium abundance with the increase in  $M_{\Sigma}$ near $z\simeq1$. With the increase in $M_{\Sigma}$, the rate of gauge boson-mediated annihilations for $\Sigma$ decreases. This makes $\Sigma$ to deviate more from its equilibrium abundance. It can, also, be seen that with the increase in $M_{N,\Sigma}$ the $B-L$ asymmetry increases. This is due to the suppression of inverse decay washouts with the increase in $M_{N,\Sigma}$. In Fig. \ref{bl2} we have shown the evolution of $n_{N}$, $n_{\Sigma}$ and $n_{B-L}$ with $z=M_{N}/T$ for different values of $\lambda_{5}$ in two moderately hierarchical heavy fermions case. With the decrease in $\lambda_{5}$ the Yukawa couplings for $N,\Sigma$ increases through the CI parameterisation of Eqn. (\ref{eq:CI}). Due to the increase in Yukawa coupling the decay and inverse decay rates increase resulting in $n_{N,\Sigma}$ to remain closer to their equilibrium abundance. Similarly from the lower panel plot of Fig. \ref{bl2} it can be seen that with the decrease in $\lambda_{5}$ the $B-L$ asymmetry first increases, however, after a certain small value of $\lambda_{5}$ it starts decreasing with further decrease in $\lambda_{5}$. For the small value of $\lambda_{5}$ the Yukawa coupling becomes large enough to increase the CP asymmetry resulting in an increase in $B-L$ asymmetry. However, beyond a certain large value of the Yukawa coupling the washouts increase which dominates over the increase in CP asymmetry. As a result, the $B-L$ asymmetry decreases with the decrease in $\lambda_{5}$.

\noindent In the presence of three fermions ($N_{1}$,$N_{2}$,$\Sigma$), the lightest active neutrino mass eigenstate becomes non-zero ($i.e.$ $m_{1}\neq 0$). Moreover, the selection of the \( R \) matrix is conducted to ensure that within the \( h \) matrix, the Yukawa couplings of \( N_1 \) manifest a proportional relationship with \( m_{1} \) characterized by \( \sqrt{m_1} \), as illustrated in Eqns. (\ref{eqn59}) and (\ref{eqn60}) of Appendix \ref{appendix2}. As $m_{1}$ approaches vanishingly small values, the Yukawa coupling associated with the lightest fermion, $N_{1}$, concurrently diminishes to an extent where a weak washout leptogenesis scenario ensues ($K_{N_{1}}<1$). For $m_{1}$ $\lesssim 10^{-5}$ eV, the effects of washouts become negligible. Furthermore, a specific mass hierarchy is chosen among the heavy fermions to ensure that out-of-equilibrium decay of both $N_1$ and $\Sigma$ can contribute to the final lepton asymmetry, i.e., $M_{N_{1}} \lesssim M_{\Sigma} \ll M_{N_{2}}$. Subsequently, solving the coupled Boltzmann Eqns. (\ref{bol21}), (\ref{bol22}), and (\ref{bol23}) results in decrease of the leptogenesis scale from $10^{10}$ GeV to $\rm TeV$ scale. In Fig. \ref{bl3} we have  shown the evolution of $n_{N_{1}},n_{\Sigma}$ and $n_{B-L}$ with $z=M_{N_{1}}/T$ for different values of $\lambda_{5}$ in three heavy fermions case. Similar behaviour of the $n_{N_{1}}$, $n_{\Sigma}$ and $n_{B-L}$ can be seen with the change in $\lambda_{5}$ as in the case of the two heavy fermions. In Fig. \ref{bl4} we show the evolution of $n_{N_{1}}$, $n_{\Sigma}$ and $n_{B-L}$ with $z=M_{N_{1}}/T$ for different values of $M_{N_{1}}$ with $M_{\Sigma}=10^{0.5}M_{N_{1}}$ for fixed benchmark values of other relevant parameters. Similar to two fermion cases, we can see the increase in asymmetry with the increase in $M_{N_{1}}$ and $M_{\Sigma}$. Furthermore, previous investigations indicate that the lightest heavy fermion is the sole contributor to the $B-L$ asymmetry, while heavier fermions are disregarded due to their substantial washout effects. However, our study suggests that when we choose moderately hierarchical heavy fermion masses, the heavier fermions can significantly contribute to the $B-L$ asymmetry.\\
The evolution of the $B-L$ asymmetry for the $N_1 +\Sigma$ (red dashed line) and $N_1$ only (blue line) is depicted in Fig. (\ref{compare}). For later case, the $B-L$ asymmetry remains very low compared to the required value of $n_{B-L}$. However, considering the hierarchy of the heavy fermions ($M_{N_{1}}\lesssim M_{\Sigma}<M_{N_{2}}$), the triplet fermion contributes significantly to the final $B-L$ asymmetry which is, also, evident from the Fig. (\ref{compare}). Further, it is to be noted that the scalar triplet ($\Omega)$ acquires VEV at the electroweak scale, which results in the mixing of heavy fermions. At TeV scale, the singlet and triplet fermions don’t mix. Therefore, the scalar triplet does not affect leptogenesis.
\noindent Despite the lack of signatures of $\Sigma$ beyond 1 TeV at LHC, it is still feasible to explore the triplet fermions in final states that involve multiple leptons and fat-jets. These final states are more refined compared to the typical LHC searches and can be extended up to 10 TeV\cite{Ashanujjaman:2021zrh,Li:2022kkc}. At 95$\%$ confidence level, the ATLAS experiment at LHC has excluded the triplet fermion with mass below 790 GeV\cite{ATLAS:2020wop}. We have successful leptogenesis around the TeV scale. Collider experiments already constrain light triplet fermions, but future experiments may test viable parameter space. 

\noindent \textbf{Benchmark points yielding correct $n_{B-L}$:} For two heavy fermion case:\\
 $M_N=7.97\times 10^{9}$ GeV, $\lambda_{5}$=-0.1 and $M_N=6.30\times 10^{9}$ GeV, $\lambda_{5}$=-0.3.\\
For three heavy fermion case:\\
$M_{N_{1}}=1513.56$ GeV, $\lambda_{5}$=-0.1 and $M_{N_{1}}=1995.26$ GeV, $\lambda_{5}$=-0.2.

\section{Summary}\label{summary}
We examined the singlet-triplet scotogenic model (STSM) in which dark sector particles run in the loop and generate tiny neutrino masses at one loop level. The particle content of the model includes the heavy $SU(2)_{L}$ singlet $(N)$, triplet $(\Sigma)$ fermions. Scalar sector is extended by a $SU(2)_{L}$ doublet scalar $\eta$ and a hyperchargeless $SU(2)_{L}$ triplet $\Omega$. We examined the prospects of the model meeting the recent $W$-boson mass measurements of the CDF-II collaboration at tree-level through hyperchargeless scalar triplet $\Omega$. Also, we studied the phenomenology of DM where the real component of $\eta$ acts as a viable WIMP-type DM candidate.\\
Our study explored the possibility of producing baryogenesis at the TeV scale through leptogenesis in the STSM. The lepton asymmetry is generated through the out-of-equilibrium decay of both singlet (N) and triplet ($\Sigma$) fermions, with their masses chosen to be moderately degenerate. This lepton asymmetry is then converted into the baryon asymmetry via sphaleron processes that accidentally conserve $B-L$ but violate $B+L$ symmetry. In our analysis, we are examining two scenarios: one involving two heavy fermions and the other involving three, and we explicitly distinguish between the two cases. First of all, we deal with two heavy fermions, namely N and $\Sigma$, which result in the vanishing lightest neutrino mass eigenvalue ($m_{1}=0$). Both heavy fermions have a moderately hierarchical mass, and their out-of-equilibrium decay contributes to the final $B-L$ asymmetry. As light neutrino masses require a large value of the Yukawa coupling through Casas-Ibarra parameterization, the scale of leptogenesis is estimated to be around $10^{10}$ GeV, which is similar to the thermal leptogenesis in the Type-I seesaw scenario. In order to lower the scale of leptogenesis to around TeV scale, we introduce an additional heavy 
 singlet fermion and consider three fermion cases ($N_{1}$, $N_{2}$, $\Sigma$). As a result, the lightest neutrino mass now has a non-zero value ($m_{1}\neq0$). In the scotogenic model, with three heavy Majorana fermions, the scale of leptogenesis depends heavily on the value of the lightest neutrino mass ($m_{1}$). For vanishingly small values of $m_{1}$ ($\lesssim 10^{-5}$ eV), the impact of washouts is negligible. Therefore, reducing the $m_{1}$ beyond this range results in lowering the leptogenesis scale. This mechanism requires leptogenesis to end before electroweak sphaleron drop out of equilibrium. The analysis yields a minimum mass estimate of around $3$ TeV for the lightest right-handed fermion which is quite small as compared to the mass bound of $\sim 10^{9}$ GeV in standard thermal leptogenesis. Moreover, the quadratic coupling $\lambda_{5}$ plays a significant role in bridging DM and leptogenesis. Future collider experiments may offer insights into the gauge interactions of a TeV-scale triplet fermion.\\

\noindent\textbf{\Large{Acknowledgments}}
 \vspace{.3cm}\\
L. S. acknowledges the financial support provided by Central University of Himachal Pradesh. The authors, also, thank Sushant Yadav for the helpful discussions.

\appendix
\section{Relevant annihilation and coannihilation diagrams contributing to the relic abundance of $\eta^{R}$}\label{dmchanels}
\begin{figure}[H]
\centering
\begin{tikzpicture}
\begin{feynman}
\vertex at (0,0) (i1);
\vertex at (-1.5,0) (i2);
\vertex at (1.5,1) (a);
\vertex at (1.5,-1) (b);
\vertex at (-2.5,1) (c);
\vertex at (-2.5,-1) (d);

\vertex at (-2.7,-1.2) () {\(\eta^{R}\)};
\vertex at (-2.7,1.2) () {\(\eta^{I}\)};
\vertex at (1.7,1.2) () {\(\l^{-},\overline{q}\)};
\vertex at (1.7,-1.2) () {\(\l^{+},q\)};
\vertex at (-0.5,-0.2) () {\(Z\)};
\diagram*{
(i2) -- [photon] (i1), (i1) -- [fermion] (a), (b) -- [fermion] (i1), (i2) -- [charged scalar] (c),(d) -- [charged scalar] (i2)
};
\end{feynman}
\end{tikzpicture}
\hspace{0.2cm}  \begin{tikzpicture}
\begin{feynman}
\vertex at (0,0) (i1);
\vertex at (-1.5,0) (i2);
\vertex at (1.5,1) (a);
\vertex at (1.5,-1) (b);
\vertex at (-2.5,1) (c);
\vertex at (-2.5,-1) (d);

\vertex at (-2.7,-1.2) () {\(\eta^{R}\)};
\vertex at (-2.7,1.2) () {\(\eta^{I}\)};
\vertex at (1.7,1.2) () {\(\l^{-},\overline{q}\)};
\vertex at (1.7,-1.3) () {\(\l^{+},q\)};
\vertex at (-0.5,-0.3) () {\(W^{\pm}\)};
\diagram*{
(i2) -- [photon] (i1), (i1) -- [fermion] (a), (b) -- [fermion] (i1), (i2) -- [charged scalar] (c),(d) -- [charged scalar] (i2)
};
\end{feynman}
\end{tikzpicture}
\hspace{0.2cm}  \begin{tikzpicture}
\begin{feynman}
\vertex at (0,0) (i1);
\vertex at (2.2,1.3) (a);
\vertex at (2.2,-1) (b);
\vertex at (-2.2,1.3) (c);
\vertex at (-2.2,-1) (d);

\vertex at (-2.4,-1.3) () {\(\eta^{R}\)};
\vertex at (-2.4,1.5) () {\(\eta^{R}\)};
\vertex at (2.4,1.5) () {\(W^{-}\)};
\vertex at (2.4,-1.5) () {\(W^{+}\)};
\diagram*{
(i1) -- [photon] (a), (b) -- [photon] (i1), (c) -- [charged scalar] (i1),(d) -- [charged scalar] (i1)
};
\end{feynman}
\end{tikzpicture}
\hspace{0.2cm}  \begin{tikzpicture}
\begin{feynman}
\vertex at (0,0) (i1);
\vertex at (-1.5,0) (i2);
\vertex at (1.5,1) (a);
\vertex at (1.5,-1) (b);
\vertex at (-2.5,1) (c);
\vertex at (-2.5,-1) (d);

\vertex at (-2.7,-1.3) () {\(\eta^{R}\)};
\vertex at (-2.7,1.3) () {\(\eta^{I}\)};
\vertex at (1.7,1.3) () {\(W^{-}\)};
\vertex at (1.7,-1.3) () {\(W^{+}\)};
\vertex at (-0.5,-0.2) () {\(Z\)};
\diagram*{
(i2) -- [photon] (i1), (i1) -- [photon] (a), (b) -- [photon] (i1), (i2) -- [charged scalar] (c),(d) -- [charged scalar] (i2)
};
\end{feynman}
\end{tikzpicture}
\hspace{0.2cm}  \begin{tikzpicture}
\begin{feynman}
\vertex at (0,0) (i1);
\vertex at (-1.5,0) (i2);
\vertex at (1.5,1) (a);
\vertex at (1.5,-1) (b);
\vertex at (-2.5,1) (c);
\vertex at (-2.5,-1) (d);

\vertex at (-2.7,-1.3) () {\(\eta^{R}\)};
\vertex at (-2.7,1.3) () {\(\eta^{I}\)};
\vertex at (1.7,1.3) () {\(\l^{-},\overline{q}\)};
\vertex at (1.7,-1.3) () {\(\l^{+},q\)};
\vertex at (-0.5,-0.3) () {\(H_{1},H_{2}\)};
\diagram*{
(i2) -- [charged scalar] (i1), (i1) -- [fermion] (a), (b) -- [fermion] (i1), (i2) -- [charged scalar] (c),(d) -- [charged scalar] (i2)
};
\end{feynman}
\end{tikzpicture}
\hspace{0.2cm}  \begin{tikzpicture}
\begin{feynman}
\vertex at (0,0) (i1);
\vertex at (-1.5,0) (i2);
\vertex at (1.5,1) (a);
\vertex at (1.5,-1) (b);
\vertex at (-2.5,1) (c);
\vertex at (-2.5,-1) (d);

\vertex at (-2.7,-1.3) () {\(\eta^{R}\)};
\vertex at (-2.7,1.3) () {\(\eta^{I}\)};
\vertex at (1.7,1.3) () {\(\nu\)};
\vertex at (1.7,-1.3) () {\(l^{\pm}\)};
\vertex at (-0.5,-0.3) () {\(H_{1,2}^{\pm}\)};
\diagram*{
(i2) -- [charged scalar] (i1), (i1) -- [fermion] (a), (b) -- [fermion] (i1), (i2) -- [charged scalar] (c),(d) -- [charged scalar] (i2)
};
\end{feynman}
\end{tikzpicture}
\hspace{0.2cm}  \begin{tikzpicture}
\begin{feynman}
\vertex at (0,0) (a);
\vertex at (1.7,-1) (i1);
\vertex at (-1.7,-1) (i2);
\vertex at (0,1) (b);
\vertex at (1.7,2) (c);
\vertex at (-1.7,2) (d);
\vertex at (1.8,2.3) () {\(W^{\pm}\)};
\vertex at (-1.8,2.3) () {\(\eta^{I}\)};
\vertex at (1.8,-1) () {\(W^{+}\)};
\vertex at (-1.8,-1) () {\(\eta^{R}\)};
\vertex at (0.4,0.7) () {\(\eta^{\pm}\)};

\diagram*{
(i1) -- [photon] (a), (i2) -- [charged scalar] (a),
(a) -- [charged scalar] (b), (c) -- [photon] (b), (d) -- [charged scalar] (b),
};
\end{feynman}
\end{tikzpicture}
\hspace{0.2cm}
\begin{tikzpicture}
\begin{feynman}
\vertex at (0,0) (i1);
\vertex at (2.2,1.3) (a);
\vertex at (2.2,-1) (b);
\vertex at (-2.2,1.3) (c);
\vertex at (-2.2,-1) (d);

\vertex at (-2.4,-1.3) () {\(\eta^{R}\)};
\vertex at (-2.4,1.5) () {\(\eta^{R}\)};
\vertex at (2.4,1.5) () {\(Z\)};
\vertex at (2.4,-1.3) () {\(Z\)};
\diagram*{
(i1) -- [photon] (a), (b) -- [photon] (i1), (c) -- [charged scalar] (i1),(d) -- [charged scalar] (i1)
};
\end{feynman}
\end{tikzpicture}
\hspace{0.2cm}
\begin{tikzpicture}
\begin{feynman}
\vertex at (0,0) (a);
\vertex at (1.7,-1) (i1);
\vertex at (-1.7,-1) (i2);
\vertex at (0,1) (b);
\vertex at (1.7,2) (c);
\vertex at (-1.7,2) (d);
\vertex at (1.8,2.2) () {\(l^{\pm}\)};
\vertex at (-1.8,2.2) () {\(\eta^{R/I}\)};
\vertex at (1.8,-1) () {\(l^{\pm}\)};
\vertex at (-1.8,-1) () {\(\eta^{R}\)};
\vertex at (0.4,0.7) () {\(\chi^{\pm}\)};

\diagram*{
(i1) -- [fermion] (a), (i2) -- [charged scalar] (a),
(a) -- [fermion] (b), (c) -- [fermion] (b), (d) -- [charged scalar] (b),
};
\end{feynman}
\end{tikzpicture}
\begin{tikzpicture}
\begin{feynman}
\vertex at (0,0) (i1);
\vertex at (-1.5,0) (i2);
\vertex at (1.5,1) (a);
\vertex at (1.5,-1) (b);
\vertex at (-2.5,1) (c);
\vertex at (-2.5,-1) (d);

\vertex at (-2.7,-1.2) () {\(\eta^{R}\)};
\vertex at (-2.7,1.2) () {\(\eta^{R}\)};
\vertex at (1.7,1.2) () {\(W^{-}\)};
\vertex at (1.7,-1.2) () {\(W^{+}\)};
\vertex at (-0.5,-0.3) () {\(H_{1},H_{2}\)};
\diagram*{
(i2) -- [charged scalar] (i1), (i1) -- [photon] (a), (b) -- [photon] (i1), (i2) -- [charged scalar] (c),(d) -- [charged scalar] (i2)
};
\end{feynman}
\end{tikzpicture}
\hspace{0.1cm}
\begin{tikzpicture}
\begin{feynman}
\vertex at (0,0) (a);
\vertex at (1.7,-1) (i1);
\vertex at (-1.7,-1) (i2);
\vertex at (0,1) (b);
\vertex at (1.7,2) (c);
\vertex at (-1.7,2) (d);
\vertex at (1.8,2.2) () {\(\nu\)};
\vertex at (-1.8,2.2) () {\(\eta^{R/I}\)};
\vertex at (1.8,-1) () {\(\nu\)};
\vertex at (-1.8,-1) () {\(\eta^{R}\)};
\vertex at (0.4,0.7) () {\(\chi_{\sigma}\)};

\diagram*{
(i1) -- [fermion] (a), (i2) -- [charged scalar] (a),
(a) -- [fermion] (b), (c) -- [fermion] (b), (d) -- [charged scalar] (b),
};
\end{feynman}
\end{tikzpicture}
\hspace{0.1cm}
\begin{tikzpicture}
\begin{feynman}
\vertex at (0,0) (a);
\vertex at (1.7,-1) (i1);
\vertex at (-1.7,-1) (i2);
\vertex at (0,1) (b);
\vertex at (1.7,2) (c);
\vertex at (-1.7,2) (d);
\vertex at (1.8,2.2) () {\(H_{1},H_{2}\)};
\vertex at (-1.8,2.2) () {\(\eta^{R}\)};
\vertex at (1.8,-1) () {\(H_{1},H_{2}\)};
\vertex at (-1.8,-1) () {\(\eta^{R}\)};
\vertex at (0.4,0.7) () {\(H_{1}\)};

\diagram*{
(i1) -- [charged scalar] (a), (i2) -- [charged scalar] (a),
(a) -- [charged scalar] (b), (c) -- [charged scalar] (b), (d) -- [charged scalar] (b),
};
\end{feynman}
\end{tikzpicture}
\hspace{0.2cm}
\begin{tikzpicture}
\begin{feynman}
\vertex at (0,0) (a);
\vertex at (1.7,-1) (i1);
\vertex at (-1.7,-1) (i2);
\vertex at (0,1.2) (b);
\vertex at (1.7,2.2) (c);
\vertex at (-1.7,2.2) (d);
\vertex at (1.8,2.4) () {\(\nu\)};
\vertex at (-1.8,2.4) () {\(\eta^{\pm}\)};
\vertex at (1.8,-1.1) () {\(l^{\pm}\)};
\vertex at (-1.8,-1.1) () {\(\eta^{R}\)};
\vertex at (0.4,0.7) () {\(\chi^{\pm}\)};

\diagram*{
(i1) -- [fermion] (a), (i2) -- [charged scalar] (a),
(a) -- [fermion] (b), (c) -- [fermion] (b), (d) -- [charged scalar] (b),
};
\end{feynman}
\end{tikzpicture}
\hspace{0.1cm}
\begin{tikzpicture}
\begin{feynman}
\vertex at (0,0) (a);
\vertex at (1.7,-1) (i1);
\vertex at (-1.7,-1) (i2);
\vertex at (0,1.5) (b);
\vertex at (1.7,2.5) (c);
\vertex at (-1.7,2.5) (d);
\vertex at (1.8,2.7) () {\(W^{-}\)};
\vertex at (-1.8,2.7) () {\(\eta^{R}\)};
\vertex at (1.8,-1.3) () {\(W^{+}\)};
\vertex at (-1.8,-1.3) () {\(\eta^{R}\)};
\vertex at (0.4,1) () {\(\eta^{\pm}\)};

\diagram*{
(i1) -- [photon] (a), (i2) -- [photon] (a),
(a) -- [charged scalar] (b), (c) -- [charged scalar] (b), (d) -- [charged scalar] (b),
};
\end{feynman}
\end{tikzpicture}
\hspace{0.2cm}
\begin{tikzpicture}
\begin{feynman}
\vertex at (0,0) (a);
\vertex at (1.7,-1) (i1);
\vertex at (-1.7,-1) (i2);
\vertex at (0,1.5) (b);
\vertex at (1.7,2.5) (c);
\vertex at (-1.7,2.5) (d);
\vertex at (1.8,2.7) () {\(l^{\pm}\)};
\vertex at (-1.8,2.7) () {\(\eta^{\pm}\)};
\vertex at (1.8,-1.3) () {\(\nu\)};
\vertex at (-1.8,-1.3) () {\(\eta^{R}\)};
\vertex at (0.4,1) () {\(\chi_{\sigma}\)};

\diagram*{
(i1) -- [fermion] (a), (i2) -- [charged scalar] (a),
(a) -- [fermion] (b), (c) -- [fermion] (b), (d) -- [charged scalar] (b),
};
\end{feynman}
\end{tikzpicture}
\hspace{0.2cm}
\begin{tikzpicture}
\begin{feynman}
\vertex at (0,0) (i1);
\vertex at (-1.5,0) (i2);
\vertex at (1.5,1.5) (a);
\vertex at (1.5,-1.5) (b);
\vertex at (-2.5,1.5) (c);
\vertex at (-2.5,-1.5) (d);

\vertex at (-2.7,-1.7) () {\(\eta^{R}\)};
\vertex at (-2.7,1.7) () {\(\eta^{R}\)};
\vertex at (1.7,1.7) () {\(H_{1},H_{2}\)};
\vertex at (1.7,-1.7) () {\(H_{1},H_{2}\)};
\vertex at (-0.5,-0.2) () {\(H_{1}\)};
\diagram*{
(i2) -- [photon] (i1), (i1) -- [charged scalar] (a), (b) -- [fermion] (i1), (i2) -- [charged scalar] (c),(d) -- [charged scalar] (i2)
};
\end{feynman}
\end{tikzpicture}
\end{figure}

\section{Two heavy fermion case}\label{appendix1}

When we consider two heavy fermions $N$ and $\Sigma$ the lightest active neutrino becomes massless and the CI parametrisation gives the Yukawa couplings for the heavy fermions

\begin{eqnarray}
    h=U^{*} \sqrt{\widetilde{M}} R \sqrt{\Lambda}^{-1}.
\end{eqnarray}

\noindent Where $\widetilde{M}=diag(m_{1},m_{2},m_{3})$ is the active neutrino mass matrix.
To generate a non-zero CP asymmetry $\epsilon^{N}$ and $\epsilon^{\Sigma}$ the following $R$ matrix is chosen 

\begin{equation}
    R= \begin{pmatrix}
        0 & 0 \\ \cos\theta & \sin \theta \\ -\sin \theta & \cos \theta 
    \end{pmatrix}.
\end{equation}

\noindent The explicit form of the Yukawa couplings is then found to be 

\begin{eqnarray}
    h=\begin{pmatrix}
        \Lambda_{1}^{-1/2}(\sqrt{m_{2}} \cos\theta u_{21}^{*}+ \sqrt{m_{3}} \sin\theta u_{31}^{*}  ) &  \Lambda_{2}^{-1/2}(-\sqrt{m_{2}} \sin\theta u_{21}^{*}+ \sqrt{m_{3}} \cos\theta u_{31}^{*}  ) \\
         \Lambda_{1}^{-1/2}(\sqrt{m_{2}} \cos\theta u_{22}^{*}+ \sqrt{m_{3}} \sin\theta u_{32}^{*}  ) &\Lambda_{2}^{-1/2}(-\sqrt{m_{2}} \sin\theta u_{22}^{*}+ \sqrt{m_{3}} \cos\theta u_{32}^{*}  ) \\
         \Lambda_{1}^{-1/2}(\sqrt{m_{2}} \cos\theta u_{23}^{*}+ \sqrt{m_{3}} \sin\theta u_{33}^{*}  ) & \Lambda_{2}^{-1/2}(-\sqrt{m_{2}} \sin\theta u_{23}^{*}+ \sqrt{m_{3}} \cos\theta u_{33}^{*}  )
    \end{pmatrix}.
    \label{eq:Yuk}
\end{eqnarray}

\noindent From Eqn. (\ref{eq:Yuk}) it can be seen that the Yukawa couplings are dependent on $m_{2}$ and $m_{3}$. With the normal ordering of light neutrino mass, we don't have freedom over $m_{2}$ and $m_{3}$ to make the Yukawa coupling smaller irrespective of the value of $\theta$. Here $u_{ij}$s are the elements of the PMNS matrix. The only way to make the Yukawa couplings smaller is to increase the scalar coupling $\lambda_{5}$. Since $\lambda_{5}$ has a perturbative limit of $4\pi$, the Yukawa couplings can not be made arbitrarily small by increasing the value of $\lambda_{5}$. Therefore we are always in a strong washout region with two heavy fermions.

\section{Three heavy fermion case}\label{appendix2}

With the presence of another heavy fermion ($N_{1}$, $\Sigma$, $N_{2}$) the minimal $R$ matrix to generate a non-zero CP asymmetry become

\begin{equation}\label{eqn59}
    R= \begin{pmatrix}
        1&0 & 0 \\ 0&\cos\theta & \sin \theta \\ 0&-\sin \theta & \cos \theta 
    \end{pmatrix},
\end{equation}

\noindent and the Yukawa matrix can be found to be

\begin{eqnarray}\label{eqn60}
    h=\begin{pmatrix}
       \sqrt{m_{1}}\Lambda_{1}^{-1/2} u^{*}_{11} & \Lambda_{1}^{-1/2}(\sqrt{m_{2}} \cos\theta u_{21}^{*}+ \sqrt{m_{3}} \sin\theta u_{31}^{*}  ) &  \Lambda_{2}^{-1/2}(-\sqrt{m_{2}} \sin\theta u_{21}^{*}+ \sqrt{m_{3}} \cos\theta u_{31}^{*}  ) \\
     \sqrt{m_{1}}\Lambda_{1}^{-1/2} u^{*}_{12} &    \Lambda_{1}^{-1/2}(\sqrt{m_{2}} \cos\theta u_{22}^{*}+ \sqrt{m_{3}} \sin\theta u_{32}^{*}  ) &\Lambda_{2}^{-1/2}(-\sqrt{m_{2}} \sin\theta u_{22}^{*}+ \sqrt{m_{3}} \cos\theta u_{32}^{*}  ) \\
     \sqrt{m_{1}}\Lambda_{1}^{-1/2} u^{*}_{13} &    \Lambda_{1}^{-1/2}(\sqrt{m_{2}} \cos\theta u_{23}^{*}+ \sqrt{m_{3}} \sin\theta u_{33}^{*}  ) & \Lambda_{2}^{-1/2}(-\sqrt{m_{2}} \sin\theta u_{23}^{*}+ \sqrt{m_{3}} \cos\theta u_{33}^{*}  )
    \end{pmatrix}.
    \label{eq:Yuk2}
\end{eqnarray}

\noindent From Eqn. (\ref{eq:Yuk2}) it can be seen that the Yukawa couplings for the lightest of the three heavy fermions ($N_{1}$) can be made small by choosing the vanishingly small lightest active neutrino mass $m_{1}$. This makes the decay and the inverse decay widths smaller, resulting in a weak washout leptogenesis scenario. Therefore, the leptogenesis scale can be significantly lower compared to the two generations of heavy fermions.

\end{document}